\documentclass[journal,draftclsnofoot,onecolumn,12pt]{IEEEtran}

\normalsize

\ifCLASSINFOpdf
  \usepackage[pdftex]{graphicx}
\else
\fi

\usepackage[cmex10]{amsmath}
\usepackage{amssymb}
\usepackage{amsthm}
\theoremstyle{plain}
\newtheorem{theorem}{Theorem}
\theoremstyle{definition}
\newtheorem{definition}{Definition}

\theoremstyle{plain}

\newtheorem{lemma}{Lemma}

\theoremstyle{remark}
\newtheorem{remark}{Remark}

\allowdisplaybreaks[4]

\usepackage[margin=0.3in]{caption}
\usepackage[font=footnotesize]{subfig}
\usepackage[bookmarks=true, bookmarksnumbered=true]{hyperref}

\newcommand{\op}[1]{\operatorname{#1}}
\newcommand\numberthis{\addtocounter{equation}{1}\tag{\theequation}}

\usepackage[framemethod=TikZ]{mdframed}

\mdfdefinestyle{exampledefault}{%
rightline=true,innerleftmargin=10,innerrightmargin=10,
frametitlerule=true,
frametitlerulewidth=1pt,linecolor=white,
tikzsetting={draw=black,line width=.5pt,dashed,dash pattern= on 1pt off 3pt}}

\mdfdefinestyle{solidbox}{%
topline=true,
bottomline=true,
leftline=true,
rightline=true,
innerleftmargin=10,innerrightmargin=10,
innerbottommargin=5,
frametitlerule=false,
frametitlerulewidth=1pt}

\usepackage{varwidth}
\usepackage{algpseudocode}

\begin{document}
%
\title{Fair Scheduling Policies Exploiting Multiuser Diversity in Cellular Systems with Device-to-Device Communications}
%
%
%

\author{PhuongBang~Nguyen,~\IEEEmembership{Student Member,~IEEE,}
        and~Bhaskar~Rao,~\IEEEmembership{Fellow,~IEEE}
\thanks{This research was supported by the National Science Foundation Grant No. CCF-1115645, the UC Discovery Grant com-212531, the Broadcom Foundation Gift, the Ericsson Chair funds, and Center of Excellence for Telecom Applications, King Abdulaziz City for Science and Technology, Riyadh, Saudi Arabia.

PhuongBang Nguyen and Bhaskar Rao are with the Dept. of Electrical and Computer Engineering, University of California, San Diego.
\vspace{-0.1in}}
}

\maketitle

\pagestyle{empty}
\thispagestyle{empty}

\vspace{-0.8in}

\begin{abstract}
\vspace{-0.15in}
We consider the resource allocation problem in cellular networks which support Device-to-Device Communications (D2D).  For systems that enable D2D via only orthogonal resource sharing, we propose and analyze two resource allocation policies that guarantee access fairness among all users, while taking advantage of multi-user diversity and local D2D communications, to provide marked improvements over existing cellular-only policies.  The first policy, the \emph{Cellular Fairness Scheduling (CFS)} Policy, provides the simplest D2D extension to existing cellular systems, while the second policy, the \emph{D2D Fairness Scheduling (DFS)} Policy, harnesses maximal performance from D2D-enabled systems under the orthogonal sharing setting.  For even higher spectral efficiency, cellular systems with D2D can schedule the same frequency resource for more than one D2D pairs.  Under this non-orthogonal sharing environment, we propose a novel group scheduling policy, the \emph{Group Fairness Scheduling (GFS)} Policy, that exploits both spatial frequency reuse and multiuser diversity in order to deliver dramatic improvements to system performance with perfect fairness among the users, regardless of whether they are cellular or D2D users.
\end{abstract}

\vspace{-0.2in}

\begin{IEEEkeywords}
\vspace{-0.15in}
D2D, Device-to-Device Communications, Fairness, Orthogonal Sharing, Non-orthogonal Sharing, Group Scheduling, Uniform Performance Index, Group Allocation, Multiuser Diversity
\end{IEEEkeywords}

%
\pagebreak

\section{Introduction} \label{sec:intro}
%
%
%
%



The recent mobile computing revolution has brought about the explosive growth of smart devices.  In order to satisfy the communication needs for all these new devices under the scarcity of the available radio frequency (RF) bandwidth, the wireless cellular systems must employ many new advanced technologies to maximize spectral efficiency and reuse.  Device-to-Device communications (D2D) is one such enabling technologies for next generation wireless systems.  It is one of the main study items in LTE releases 12 and 13 \cite{LinAndGhoRat14} with many important identified use cases including Public Safety Broadband Networks, Commercial Proximity Services, and Network Offloading, among others \cite{Doppler2009}, \cite{Fodor2012}.  Unlike existing cellular communications where data exchanged between two connected devices must be relayed by the base station, the two devices in a D2D pair can send data directly to each other.  The D2D devices, however, remain under the base station's control for all administrative operations such as resource allocation, power control, and so on, to enable centralized resource planning and scheduling.  D2D can deliver very high bit rate thanks to the short distance between the two pairing devices.  In addition, the low D2D transmit power causes interference in only a small neighborhood, making it possible to further enhance the spectral efficiency by sharing the same spectrum among a group of D2D pairs.


D2D communications, however, present several unique challenges: new hardware enhancements for direct communications, peer and service discovery, interference management, resource allocation and scheduling.  In this paper, we address the resource allocation and scheduling problem as it holds the key to both system improvements and user experience.  Many resource allocations and scheduling algorithms exist for traditional cellular systems such as round-robin, opportunistic round-robin \cite{Patil2009}, max weight scheduling \cite{Andrews2004}, proportional fairness (PF) algorithm \cite{Kelly1998}, \cite{Jalali2000}, \cite{MoWalrand2000}, \cite{Viswanath2002}, \cite{Andrews2005}, cumulative distribution function (CDF) based scheduling \cite{Park2005}, max rate algorithm \cite{Knopp1995}, \cite{Tsybakov2002}, resource-constrained opportunistic scheduling \cite{Liu2001}, and more. However, being designed for cellular communications, existing scheduling schemes can not directly address scheduling requirements of D2D systems and must be extended appropriately, especially to handle the sharing of a single resource among multiple pairs.  Many authors choose to extend the popular proportional fairness scheme to cover the D2D environment \cite{Lee2013}, \cite{FengJunyi2013}, \cite{Hung-Hsiang2013}.  While proportional fair schedulers are known for good performance, it has been shown that CDF-based schedulers \cite{Park2005} can perform similarly or even better in a number of scenarios \cite{PatilPhdThesis2006}, \cite{NguyenA2014}.  In addition, CDF-based methods generally provide much better analytical tractability.  Therefore in this work, we design our D2D scheduling policies based on the CDF scheduling framework to take advantages of both its high performance and tractability.  


A number of existing works discuss the resource sharing issue in the D2D settings.  In \cite{Lee2013}, the authors propose sharing a low interference cellular link with a D2D pair and using the D2D PF metrics for resource allocation.  In \cite{FengJunyi2013}, the author proposes sharing a resource block between a cellular user and a D2D pair using a modified PF metric to take into account both cellular and D2D rates.  In \cite{Asadi2003d2d}, the authors propose a group scheduling policy that makes use of a secondary radio interface such as WiFi. In \cite{MinLee2011} and \cite{Wang2012}, the authors propose a non-orthogonal resource sharing scheme between the cellular users and a D2D pair by keeping the cellular users outside of an interference area.  In \cite{Yu2011}, the resource sharing between a cellular user and a D2D pair is discussed and the sum rate of the two links is maximized under a simple power control scheme.  In \cite{Xu2012}, the authors use second-price auction strategy to assign shared resources to D2D pairs. In these existing D2D-specific schemes, however, there is a lack of a tractable analytical framework to quantify the system performance improvements as well as user access fairness when D2D is enabled.  In the traditional resource allocation framework, the fairness concept has been studied extensively in the literature \cite{Kelly1998}, \cite{MoWalrand2000}, \cite{Jain1984}, \cite{Elliott2002}, \cite{Yang2007}.  Nevertheless, existing fairness measures designed for orthogonal sharing systems, where no two users can have simultaneous access to the same shared resource, do not work well without modification in capturing the user performance in non-orthogonal sharing settings such as cellular systems with D2D sharing groups.  In this work, we introduce a new performance metric that can address this issue and enable tractable formulation of the system optimization for both performance and fairness.

\textbf{\emph{Contributions.}} Our main contributions are as follows
\begin{itemize}
  \item We provide further analytical and numerical results to illustrate the performance gains for orthogonal scheduling policies introduced in our prior work \cite{Nguyen2013}.
  \item We introduce a new performance measure that can capture individual user performance in any diverse environment with different user probability density distributions.  This measure is well-suited for systems with D2D or any system under non-orthogonal sharing settings.
  \item We propose a novel group scheduling policy that takes advantage of multiuser diversity and spectrum reuse to provide excellent performance with perfect fairness for users in D2D non-orthogonal group settings and any general system with group-based resource allocation.
\end{itemize}

\textbf{\emph{Paper Organization.}} Section \ref{sec:system_model} presents the system model and scheduling background.  Section \ref{sec:ortho_d2d} presents two orthogonal scheduling policies for D2D.  In section \ref{sec:non_ortho_d2d_and_upi} we discuss the non-orthogonal D2D grouping problem and propose our group scheduling policy.  Section \ref{sec:conclusion} contains our conclusion.  Finally, the appendices contain the proofs for the included theorems.

\textbf{\emph{Notations.}} Unless otherwise noted, in general, we use capital letter $S$ to denote the user SNR random variable, $U, V$ for the CDF transformed variables, $F(.)$ for cumulative distribution function (CDF), $f(.)$ for probability density function (PDF), $\mathbf{1}_{\{\mathcal{E}\}}$ as an indicator function of event $\mathcal{E}$, which is $1$ when $\mathcal{E}$ is true, and $0$ otherwise.  Notation $[X]^{+} = X \cdot \mathbf{1}_{\{X \ge 0\}} $.  The superscripted star symbol ($^{*}$) denotes the selected condition.  Subscript letter $c$ identifies cellular users and $d$ for D2D users.  For example, $S_{k,c}$ denotes the SNR for cellular user $k$ and $S_{i, d}$ for D2D user $i$.


\section{System Model and CDF Scheduling Background} \label{sec:system_model}

\subsection{System Model and Assumptions}

We consider a single-cell cellular system with K users, $K = K_1 + 2K_2$, where $K_1$ is the number of cellular users, $K_2$ is the number of D2D pairs.  For simplicity, it is assumed that all users and the base station have a single omni-directional antenna.  All users transmit and receive in synchronized time slots.  All users feed back perfect instantaneous channel state information (CSI) for all frequency resources.  Beside the regular cellular CSI, D2D users also report the CSI's of the D2D paths when required.  We assume that users always have data to transmit and there are no service delay constraints.  In addition, we assume no interference in the orthogonal scenarios (section \ref{sec:ortho_d2d}) and negligible interference for the non-orthogonal scenario due to proper grouping of spatially separated D2D pairs (section \ref{sec:non_ortho_d2d_and_upi}).

\subsection{CDF Scheduling Background} \label{subsec:cdf_background}

Since our scheduling policies are based on CDF scheduling \cite{Park2005}, we present a quick review on its concepts to lay the ground work for subsequent discussions.  Consider a traditional cellular system where each user $k$ feeds back its channel SNR, $s_{k,c}$, which is a sample of the SNR random variable $S_{k,c}$, to the base station.  Let $u_k = F_{S_{k,c}}(s_{k,c})$, where $F_{S_{k,c}}(s)$ is the CDF of $S_{k,c}$.  The users are selected according to: $k^{*} = \underset{k}{\operatorname{arg \, max}} \, u_k^{1/w_k}$, which gives access probability of $w_k$ to user $k$, where $\sum_{k=1}^{K} w_k = 1$.  The transformed random variables $U_k = F_{S_{k,c}}(S_{k,c})$ are i.i.d and uniformly distributed in $[0, 1]$.  Thus, the user selection is \emph{independent} of the specific user channel distributions.  The values of $U_k$'s reflect how good the users' current channels are relative to \emph{their own} channel conditions.  This policy leads to a very rigorous notion of fairness where the user with the least chance to improve is served \cite{Park2005}.  In addition, this use of i.i.d uniform random variables enhances mathematical tractability and enables system performance analysis \cite{Huang2013b}.  Motivated by the fact that CDF scheduling selects users based on their CDF-mapped values, we introduce the following performance quantity, the \emph{Uniform Performance Index} (UPI) as a measure of user performance for multiuser scheduling policies.  Consider a multiuser system with of K users competing for a single common resource.  Each user $i$ is associated with a random variable $X_i$.  The set of random variables $\{X_1, X_2, \dots, X_K\}$, are independent, with possibly different density distributions.  The resource allocation problem is one of granting access to the common resource to a single user at every selection instance based on some function of $\{X_1, X_2, \dots, X_K\}$. Let $F_{X_i}(x)$ be the CDF of random variable $X_i$.  Let $U_i \triangleq F_{X_i}(X_i)$ be the CDF-transformed random variable for $X_i$.

\begin{definition} \label{def:upi}
The Uniform Performance Index (UPI) with respect to $X$ for user $i$ is
    \begin{flalign*}
    \op{UPI}_i^{(X)} = 2 \mathbf{E} \{ U_i^* \}, \text { where } U_i^* = U_i \times \mathbf{1}_{\{\text{user i selected}\}}. 
    \end{flalign*}
\end{definition}


The sum of the UPI's for all the users in the system has been considered in \cite{Patil2009}, where it is termed "system opportunism". It can be seen that the probability of access and the \emph{relative} value of $X_i$ (i.e., $U_i$) when the access is granted are both included in the UPI measure.  Thus, this measure captures the average individual user performance in the system.  Furthermore, the fact that the UPI is independent of the user's own distribution as well as other users' distributions makes it a good measure to quantify the performance and fairness of different resource allocation schemes across different environments.  Applying this new measure to the CDF scheduling scheme, we have the following theorem on its optimality: 

\begin{theorem}
\label{theorem:cdf_opt}
The Basic CDF Scheduling (BCS) policy introduced in \cite{Park2005} is \textbf{max-min optimal} with respect to user individual UPI metric.
\end{theorem}

Theorem \ref{theorem:cdf_opt} summarizes the salient properties of BCS: fairness via the max-min operation, and high performance via the UPI optimality.  Note that BCS is fair in both UPI and temporal senses.  All our subsequent scheduling policies are designed to achieve these fairness and optimality properties.  User performance under BCS is stated in theorem \ref{theorem:cdf_snr}.

\begin{theorem}
\label{theorem:cdf_snr}
Under the BCS policy, the CDF of the SNR of a user when selected is given by
    \begin{gather}
    F_{S^{*}_{k,c}}(s) = \op{Pr}[S^{*}_{k,c} < s] = \left[F_{S_{k,c}}(s) \right]^{K}, \, \forall s \ge 0, \label{eqn:snr_cdf}
    \end{gather}
    where $S^{*}_{k,c}$ is the SNR of user $k$ when selected, $S_{k,c}$ is the overall SNR for user $k$.
\end{theorem}


Theorem \ref{theorem:cdf_snr} illustrates the multiuser diversity gain achieved by BCS.  From (\ref{eqn:snr_cdf}), $F_{S^{*}_{k,c}}(s) = \left[F_{S_{k,c}}(s) \right]^{K} \le F_{S_{k,c}}(s), \forall s \ge 0$.  Note that $F_{S_{k,c}}(s)$ corresponds to the CDF of the user SNR in a round-robin policy.  This means that for any $s$, the user SNR has lower probability to be smaller than $s$ under the BCS policy than under the round-robin policy.  In other words, the probability for SNR to be higher than $s$ is larger under the BCS policy than under the round-robin one.  For a large $K$, this gain is substantial, leading to a much high average SNR and throughput.

\section{Orthogonal D2D Scheduling} \label{sec:ortho_d2d}
In this section we introduce two \emph{orthogonal} scheduling policies where for each frequency resource, only one user is given access to it in any time slot. These are simple applications of the BCS policy to the D2D environment to allow us to quantify D2D benefits analytically.


%

\subsection{Cellular Fairness Scheduling}
One of the most difficult tasks in operating D2D is the collection of D2D CSI. Since D2D direct channels are typically in good conditions due to their short distance, the role of D2D CSI is less crucial than that for cellular communications.  Thus, one simple way to enable D2D support is to do so without D2D CSI as in the following \emph{Cellular Fairness} scheduling (CFS) policy (table \ref{tab:cellular_fairness}).  Under this policy, D2D users do not participate in the channel-based user selection.  Only cellular users compete with each other, leading to an improvement in cellular user performance.  Theorems \ref{theorem:cell_fairness_prob} and \ref{theorem:cell_fairness_opt} below discuss the fairness and optimality properties of this policy.  

\begin{table}[!ht]
    \centering
    \caption{Cellular Fairness Scheduling (CFS) Policy}
    \begin{tabular}{ | p{6.3in} |}
    \hline
    For each uplink or downlink frequency resource
      \begin{enumerate}
        \item Select a cellular user according to: $k^{*} = \underset{k \in \mathcal{K}_c}{\operatorname{argmax}} \, u_k$, where $u_k = F_{S_{k,c}}(s_{k,c})$, $\mathcal{K}_c$ is the index set of cellular users.
        \item Grant access to cellular user $k^{*}$ if $u_{k^{*}} \ge u^{th}$, where $u^{th} = \left[(K-K_1)/K\right]^{1/K_1}$.
        \item If $u_{k^{*}} < u^{th}$, grant access to a randomly selected D2D user.
      \end{enumerate}
      \vspace{-10pt}
    \\ \hline
    \end{tabular}
    \label{tab:cellular_fairness}
    \vspace{-12pt}
\end{table}

\begin{theorem}
\label{theorem:cell_fairness_prob}
The CFS policy is temporally fair for all users.  That is, each user gets probability of access of $1/K$.
\end{theorem}


\begin{theorem}
\label{theorem:cell_fairness_opt}
Among all policies with the temporal fairness constraint for all users, the CFS policy is \textbf{max-min optimal} with respect to the UPI metric for cellular users.
\end{theorem}

It can be shown that $\op{UPI}^{(CFS)} \ge \op{UPI}^{(BCS)}$ with equality only when there is no D2D user ($K_1 = K$).  This gain for cellular users come from the omission of the D2D users from the selection of the policy.  The user performance under this policy is stated in theorem \ref{theorem:cell_fairness_cdf} below.


\begin{theorem}
\label{theorem:cell_fairness_cdf}
Under the CFS policy, for each frequency resource, the CDF of the SNR for a user when selected conditioned on the user spatial distribution $\pi$ is given by
    \begin{flalign}
    F_{S_{k,c}^{*} | \pi}(s) &= \left(\frac{K}{K_1} \left[F_{S_{k,c} | \pi}(s) \right]^{K_1} - \frac{2K_2}{K_1}\right)^{+} \text{ and } F_{S_{k,d}^{*} | \pi}(s) = F_{S_{k,d} | \pi}(s), \, \forall s \ge 0, \label{eqn:cell_fairness_cdfs}
    \end{flalign}
    where $F_{S_{k,c}^{*} | \pi}(s)$ and $F_{S_{k,d}^{*} | \pi}(s)$ are the CDFs for the selected SNRs of cellular and D2D users , respectively; $F_{S_{k,c} | \pi}(s)$ and $F_{S_{k,d} | \pi}(s)$ the overall CDFs of the SNRs for cellular and D2D users.
\end{theorem}


It can be shown that the cellular CDF under the CFS policy given by (\ref{eqn:cell_fairness_cdfs}) outperforms the user CDF under the BCS policy given by (\ref{eqn:snr_cdf}).  This improvement is more beneficial for cell-edge users where the average SNR is low as illustrated in figure \ref{fig:cfs_cellular_cdf_comp}.  Here the simulation is run over 100,000 channel realizations with parameters listed in table \ref{tab:sim_params} for $K_1 = 40$ cellular users and $K_2 = 30$ D2D pairs in a fixed spatial distribution.  The behavior of the cellular user farthest from the base station is plotted.  For D2D users, the D2D CDF given in (\ref{eqn:cell_fairness_cdfs}) indicates that the D2D users behave as if they were in a round-robin scheme, and their performance is strictly dependent on their direct channel statistics.  The only D2D benefit exploited by this scheme is D2D proximity gain (the short-distance communication gain).

\subsection{D2D Fairness Scheduling} \label{subsec:d2d_fairness}


\begin{figure}[!t]
\vspace{-10pt}
\centering
\begin{minipage}{.5\textwidth}
  \centering
  \includegraphics[width=3.25in, height=2in]{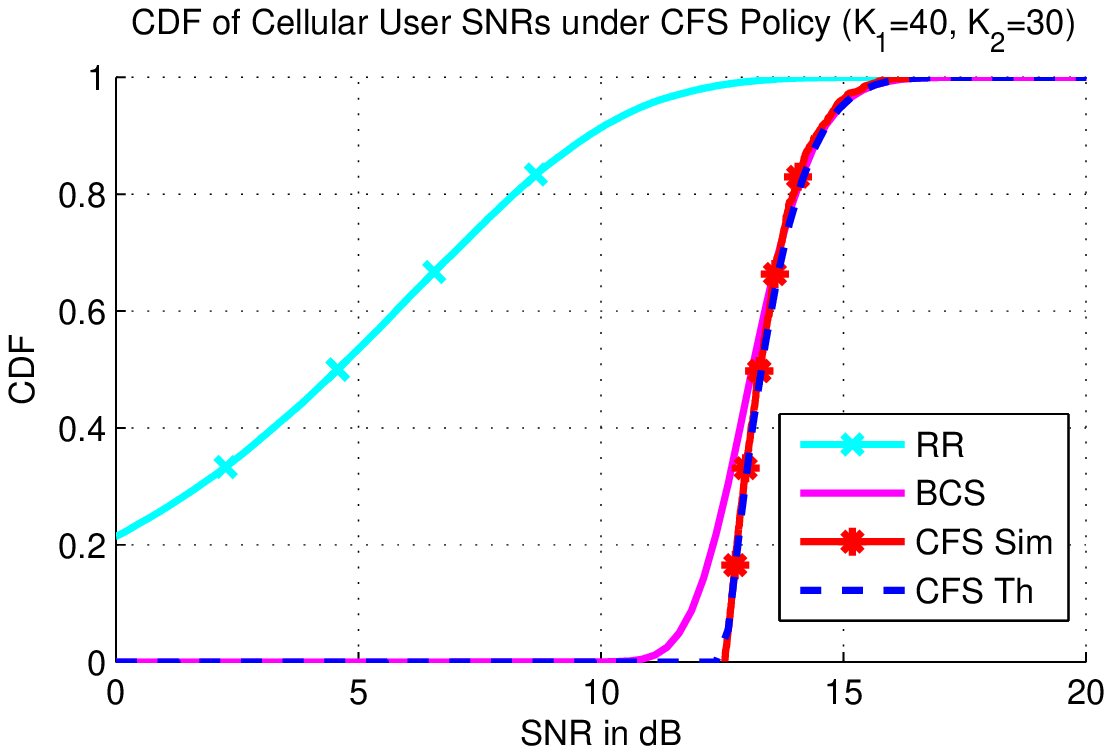}
  \captionof{figure}{CFS: cellular users get a boost in SNR}
  \label{fig:cfs_cellular_cdf_comp}
\end{minipage}%
\begin{minipage}{.5\textwidth}
  \centering
  \includegraphics[width=3.25in, height=2in]{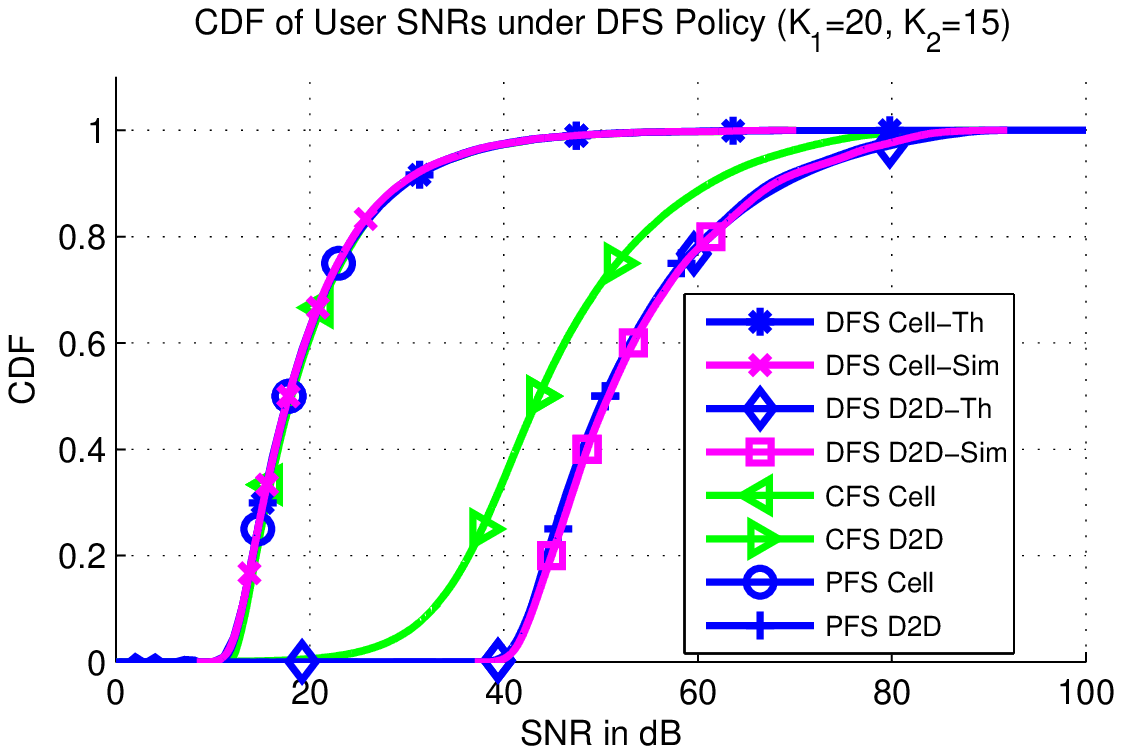}
  \captionof{figure}{DFS: D2D users have better performance than under CFS}
  \label{fig:dfs_cdf_comp}
\end{minipage}
\vspace{-20pt}
\end{figure}

\begin{table}[!t]
    \centering
    \caption{Simulation Parameters}
    \begin{tabular}{ | l | c | l | p{4cm} | l | c | l | p{5cm} |}
    \hline
    Parameter & Value & Comments         & Parameter & Value & Comments\\ \hline \hline
    Cell radius & 1000 m &               & D2D interference radius & 300 m & for spectral reuse \\ \hline
    BS antenna gain & 12 dB &            & Cellular gain constant & -31 dB & cellular $C_{c,dB}$ \\ \hline
    Mobile antenna gain & 0 dB &         & D2D gain constant & -31 dB & D2D $C_{d,dB}$           \\ \hline
    D2D min distance & 1 m & $D_{min}$   & Cellular path loss exponents & 3.5 & $\eta_c$            \\ \hline
    D2D max distance & 40 m & $D_{max}$  & D2D path loss exponents & 3 & $\eta_d$                   \\ \hline
    Noise Power & -100 dBm &             & Base station TX Power & 30 dBm & $P^{(t, dl)}$      \\ \hline
    PFS $t_c$ & 1000 & PF time constant  & D2D TX Power & 15 dBm & $P^{(t, d2d)}$              \\ \hline
    \end{tabular}
    \label{tab:sim_params}
    \vspace{-12pt}
\end{table}



The performance of D2D users can definitely improve when D2D CSI is available.  Since the two users in a D2D pair share the same direct path, they have the same fading statistic on each frequency resource.  Thus, each pair of D2D users presents \emph{only one} independent CSI to the user selection competition.  Consequently, the pair's CSI is used in every timeslot and the obtained resources is then divided between the two users according to some predefined ratio. Our scheduling approach, the \emph{D2D Fairness Scheduling} (DFS) policy, is shown in table \ref{tab:d2d_fairness}.




\begin{table}[!ht]
  \vspace{-12pt}
    \centering
    \caption{D2D Fairness Scheduling (DFS) Policy}
    \begin{tabular}{ | p{6.3in} |}
    \hline
    \begin{enumerate}

    \item For each downlink/uplink frequency resource, select a user according to:
      $k^{*} = \underset{k \in [\mathcal{K}_c \cup \mathcal{K}_{d}]}{\operatorname{argmax}} \, [U_k]^{1/w_k}$,
      where $w_k = 1/K$ for cellular users ($k \in \mathcal{K}_c$) and $w_k = 2/K$ for D2D pairs ($k \in \mathcal{K}_{d}$).
    \item For each granted D2D pair, assign the resource to one of the two users according to the predefined scheme.
    \end{enumerate} \\ \hline
    \end{tabular}
    \label{tab:d2d_fairness}
    \vspace{-12pt}
\end{table}

It is easy to see that this scheme is temporally fair.  For example, in an FDD system, for each uplink/downlink frequency resource, each cellular user receives $1/K$ access probability for every time slot, while a D2D pair receive $2/K$ for every time slot (due to the weighing property of the CDF-based selection), which averages out to $1/K$ for each D2D user per time slot.  The user performance under this policy is stated in theorem \ref{theorem:d2d_fairness_cdf} below.


\begin{theorem}
\label{theorem:d2d_fairness_cdf}
Under the DFS policy, for each frequency resource, the CDFs of the SNRs of the users when selected conditioned on the user spatial distribution $\pi$ is given by
    \begin{flalign}
    F_{S_{k,c}^{*} | \pi}(s) &= \left[F_{S_{k,c} | \pi}(s) \right]^{K} \text{ and }
    F_{S_{k,d}^{*} | \pi}(s) = \left[F_{S_{k,d} | \pi}(s)\right]^{K/2}. \label{eqn:d2d_fairness_cdfs}
    \end{flalign}
\end{theorem}
%

The cellular CDF expression (\ref{eqn:d2d_fairness_cdfs}) for cellular users is the same as (\ref{eqn:snr_cdf}).  In other words, the cellular users have the same performance as those under the BCS policy.  However, it can be seen from (\ref{eqn:d2d_fairness_cdfs}) that the D2D users greatly benefit from both D2D proximity gain (via $F_{S_{k,d}}(s)$) and multiuser diversity (via the CDF power of $K/2$). It can also be seen that this scheme benefits more from D2D communications than the CFS scheme since $F_{S_{k,d}^{*} | \pi}(s)^{DFS} = \left[F_{S_{k,d} | \pi}(s)\right]^{K/2} \le F_{S_{k,d} | \pi}(s) = F_{S_{k,d}^{*} | \pi}(s)^{CFS}$, which leads to higher a average SNR for the DFS policy.  These results are illustrated on figures  \ref{fig:dfs_cdf_comp}, which is simulated with the system parameters listed in table \ref{tab:sim_params} for $K_1 = 20$ cellular users and $K_2 = 15$ D2D pairs over 1,000,000 channel realizations with 100 user spatial distributions.  Figure \ref{fig:dfs_cdf_comp} also includes the results for Proportional Fair Scheduling (PFS) policy \cite{Jalali2000}, \cite{Viswanath2002}.  It can be seen that the DFS scheme achieves similar multiuser diversity gain to the PFS policy for both cellular and D2D users.

\subsection{Evaluation of Orthogonal Scheduling Policies} \label{subsec:policy_eval}
While results in theorems \ref{theorem:cell_fairness_cdf} and \ref{theorem:d2d_fairness_cdf} are general and true for any cellular and D2D CSI density distributions, they are conditioned on user spatial distribution.  In order to evaluate the system performance under different user conditions, \emph{unconditional} results are necessary.  However, obtaining the unconditional results for arbitrary CSI and user distributions is non-trivial.  Subsequently, we introduce CSI, power control and spatial distribution models that enable tractable analytical results.

\subsubsection{Cellular and D2D Path Models} \label{ssubsec:csi_model}
The cellular/D2D received signal at user $k$ is assumed to have the following forms
  \begin{flalign}
  y_{k,c} = \sqrt{P^{(t, ul/dl)}_{k,c} g_{k,c}} h_{k,c} x_{k,c} + n_{k,c}, \quad y_{k,d} &= \sqrt{P^{(t,d2d)}_{k,d} g_{k,d}} h_{k,d} x_{k,d} + n_{k,d}, \label{eqn:rx_signals}
  \end{flalign}
where $x_{k,c}$, $x_{k,d}$ are the normalized transmitted signals with unit power; $P^{(t, ul/dl)}_{k,c}$ is the cellular uplink (ul) or downlink (dl) transmit power; $P^{(t,d2d)}_{k,d}$ is the D2D transmit power; $g_{k,c}$, $g_{k,d}$ are the cellular and D2D long-termed average, distance-dependent path losses under Okumura-Hata model: $g_{k,c} = C_c / (d_k)^{\eta_c}$, $g_{k,d} = C_d / (d_k)^{\eta_d}$ \cite{Zander2001}.  Here $C_c$, $C_d$ are constants, dependent on antenna gain, height, and so on, $\eta_c$, $\eta_d$ are the path loss exponents, dependent on the environment, and $d_k$ is the distance between the mobile device and its transmitter. $h_{k,c}$, $h_{k,d}$ are Nakagami-m fading gains \cite{Beaulieu2005}.  $n_{k,c}$, $n_{k,d}$ are complex white Gaussian noises $n_{k,c}, n_{k,d} \sim \mathcal{CN}(0, \sigma_w)$.

\subsubsection{Power Control Model} \label{ssubsec:pwr_model}
For cellular downlink, a fixed transmit power, $P^{(t, dl)}$, is assumed for all devices.  On the uplink, it is assumed that the transmit power from user $k$, $P^{(t, ul)}_{k}$, is appropriately adjusted to compensate for the distance loss, given the device-base station distance.  This uplink/downlink power model is similar to what is used in practical systems such as LTE where a fixed power is used for the entire downlink band, and a closed loop path loss compensated power is used for each device on the uplink side.
\begin{flalign}
    P^{(r, ul)}_{k} = P^{(t, ul)}_{k} g_{k,c} = P^{(t, ul)}_{k} C_c d_k^{-\eta_c} \ge P^{(th, ul)},
\end{flalign}
where $P^{(r, ul)}_{k}$ is the uplink received power at the base station from device $k$ in cellular mode, which must meet some receiver sensitivity threshold $P^{(th, ul)}$.  A fixed transmit power, $P^{(t,d2d)}$, is assumed for all D2D direct communications.

\begin{figure}[!t]
\vspace{-10pt}
\centering
\begin{minipage}{1\textwidth}
  \centering
  \includegraphics[width=0.6\textwidth, height=1in]{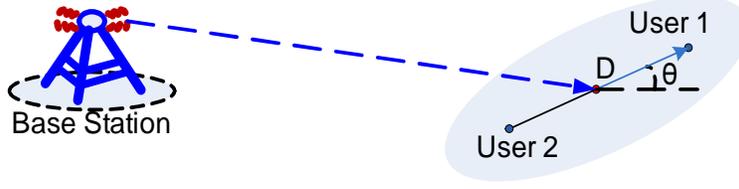}
  \captionof{figure}{D2D user distribution: pair centroids are randomly distributed, with distances uniform in {$[D_{min}, D_{max}]$}, and $\theta$ uniform in {$[0, 2 \pi)$}}
  \label{fig:d2d_dist}
\end{minipage}
\vspace{-20pt}
\end{figure}

\subsubsection{User Distribution Model}  \label{ssubsec:usr_model}
We now introduce a user distribution model that facilitates both tractable analysis and easy control of the D2D user population for simulation purposes.  The $K_1$ cellular users are distributed around the base station according to $f_{D_k}(d) = 2 d / R_B^2$, where $R_B$ is the base station radius and $D_k$ is the distance from device $k$ to the base station similar to what is done in \cite{Kang2012}.  The $K_2$ D2D-candidate pairs are distributed as shown in figure \ref{fig:d2d_dist}.  The pair centroid, which is the center of the line joining the two users, is uniformly distributed around the base station in the same fashion as the cellular users.  The vector from the centroid to one of the users has its angle $\theta$ distributed uniformly in $[0,2\pi)$.  The other user has the opposite vector from the centroid.  Different distributions can be used for the distance between the two users in a D2D pair.  For example, the Rayleigh distribution can be used to emulate the distribution of the distance between closest neighbors in a Poisson Point Process network \cite{Haenggi2005}.  For simplicity, however, we choose to use the uniform distribution in this work.  That is, the D2D distance $D$ between the two users is distributed uniformly in $[D_{min},D_{th}]$, $f_{D}(d) = 1 / (D_{th} - D_{min})$.  This particular scheme permits the number of D2D users and the direct distances of the D2D pairs to be controlled directly, allowing the evaluation of the performance across different percentages of D2D users in the system and to observe directly the effect of the D2D distance on the performance.  The randomness in both cellular and D2D user distributions makes the model richer and more realistic.

\subsubsection{Analytical Results}
Here we evaluate the unconditional CDFs for the selected SNRs.  For tractability, we let $m=1$ in the Nakagami-m fading model.  The unconditional CDFs can be evaluated by integrating over the spatial distribution.  Here we provide results for the DFS policy on \emph{downlink} frequency resources as its CDFs are more interesting.  From theorem \ref{theorem:d2d_fairness_cdf}, we have
    \begin{flalign}
    F_{S_{i,c}^{*}}^{(dl)}(s) &= \int \left[F_{S_{i,c} | \pi}(s) \right]^{K}  f_{Y_i}(y) d y \text{ and } F_{S_{k,d}^{*}}(s) = \int \left[F_{S_{k,d} | \pi}(s)\right]^{K/2} f_{X_{k,j}}(x) dx,
    \end{flalign}
where $Y_i \triangleq D_i$ is the distance between device $i$ and the base station; $f_{Y_i}(y)$ the density distribution for $Y_i$; $X_{k,j}$ the D2D direct distance between user $k$ and its pairing device $j$, and $f_{X_{k,j}}(x)$ the distribution for $X_{k,j}$.  From the CSI and power control models in subsections \ref{ssubsec:csi_model} and \ref{ssubsec:pwr_model}, after some manipulations, we get
    \begin{flalign}
    F_{S_{i,c}^{*}}^{(dl)}(s) &= \int \left[1 - \operatorname{exp} \left(-A_c s y^{\eta_c} \right) \right]^{K}  f_{Y_i}(y) dy \label{eqn:uncond_cdfa}\\
    F_{S_{k,d}^{*}}(s) &= \int \left[1 - \operatorname{exp} \left(-A_d s x^{\eta_d} \right) \right]^{\frac{K}{2}} f_{X_{k,j}}(x) dx, \label{eqn:uncond_cdfb}
    \end{flalign}
    where $A_c = \sigma_w^2 / (C_c P^{(t,dl)})$, $A_d = \sigma_w^2 / (C_d P^{(t,d2d)})$.

Expressions (\ref{eqn:uncond_cdfa}) and (\ref{eqn:uncond_cdfb}) can be evaluated given any spatial distribution for the users.  Using the spatial model in subsection \ref{ssubsec:usr_model}, we obtain the following closed form results.

\begin{theorem}
\label{theorem:dfs_uncond_cdf}
Under the DFS policy, the unconditional CDFs of selected SNRs are given by
    \begin{flalign*}
    F_{S_{c}^{*}}^{(dl)}(s) &= 1 + \sum_{i=1}^{K} {K\choose i} (-1)^{i} G_{c,i}(s) \gamma(2/\eta_c, \beta_{c,i}) \\
    F_{S_{d}^{*}}(s) &= 1 + \sum_{i=1}^{L} {K/2 \choose i} (-1)^{i} G_{d,i}(s) \left[\gamma(1/\eta_d, \beta_{d,i}) - \gamma(1/\eta_d, \alpha_{d,i}) \right],
    \end{flalign*}
    where
    \begin{flalign*}
    G_{c,i}(s) &= \frac{2}{\eta_c R_B^2 \left( i A_c s \right)^{2/\eta_c}}, \,
    G_{d,i}(s) = \frac{1}{\eta_d \Delta_D \left( i A_d s \right)^{1/\eta_d}} \\
    \beta_{c,i} &= i A_c s R_B^{\eta_c}, \, \alpha_{d,i} = i A_d s D_{min}^{\eta_d}, \, \beta_{d,i} = i A_d s D_{th}^{\eta_d}\\
    \gamma(a, x) & = \int_{0}^{x} t^{a - 1} e^{-t} dt \text{ the lower incomplete gamma function} \\
    L &= K/2 \text{ when K is multiple of 2 or } L = \infty \text{ otherwise.}
    \end{flalign*}\end{theorem}

The results in theorem \ref{theorem:dfs_uncond_cdf} are plotted together with simulation results in figure \ref{fig:dfs_cdf_comp}.  It can be seen that the theoretical results agree perfectly with simulated ones.  Again, it is clear that D2D users get better performance under the DFS policy than under the CFS policy.

\section{Non-orthogonal Sharing And Group Fairness Scheduling} \label{sec:non_ortho_d2d_and_upi}



One of the most important benefits of D2D communications is the gain in system throughput due to non-orthogonal resource sharing (spectral reuse) between multiple D2D links.  Due to the limited transmit power of each D2D user, its area of strong D2D interference is small.  Thus, instead of full interference management for all D2D pairs requiring complete knowledge of channel CSI's between all D2D users, which will be very difficult if not impractical to collect, we consider grouping together only D2D pairs that have negligible interference to each other. These are typically spatially far from each other.  There are two main tasks in a group resource allocation policy: forming the groups and allocating resources among them.  Dividing users into groups in an optimal way is generally an NP-hard (non-deterministic polynomial-time hard) problem, though heuristic solutions exist that can serve the purpose. Additional discussions on user grouping will be presented in subsection \ref{subsec:gfs_for_d2d}.  For now, we focus on the latter task of developing a novel method for allocating resources among groups.  The main challenge here is how to optimally distribute resources among groups of different sizes while maintaining fairness for all users.  Under a group sharing environment, larger groups allow higher spectral reuse at the cost of lower diversity gain for group members.  A good scheduling policy must strike a good balance between multiuser diversity and spectral reuse gains.  Under a group allocation policy, once a group wins the resource, all group members are allowed simultaneous access to it.  Simple extensions of existing scheduling policies that were designed for scheduling individual users do not work well in group sharing environments.  For example, one could pick the winner to be the group whose the best user is also the best among all groups.  This scheme, however, will lead to larger groups being granted higher probability of access compared to smaller groups, creating unfairness in the system as members of large groups now have much more access time than those of small groups.  On the other hand, giving all groups the same probability of access via priority weighting is temporally fair but results in a loss in multiuser diversity.  Even though all users now get the same probability of access, those in large groups do not get their access when their channel conditions are favorable as often as small group users.

The deficiencies of simple extensions of existing scheduling policies to group environment motivate our proposed \emph{Group Fairness Scheduling} (GFS) policy discussed subsequently.  Before we introduce this new scheduling policy, however, we need to consider the fairness issue in the system.  The parallel resource access within sharing groups provides additional "free" resources to the system beyond the available orthogonal resources.  However, this non-orthogonality renders traditional fairness measures that were designed for orthogonal access inapplicable as there are now more users than the number of resource competitors and group member users do not directly compete for resources.  As previously discussed, temporal allocation fairness is not a sufficient metric in this environment. Distribution-dependent metrics such as SNR or rate are unfair.  Hence, a new performance measure is necessary to address these challenges.  It can be seen that the \emph{Uniform Performance Index} introduced in subsection \ref{subsec:cdf_background} is very well suited.  First, the UPI is independent of the user's own distribution as well as other users' distributions as $U_i$'s are uniformly distributed.  In addition, the fact that the UPI can capture user performance for any selection scheme makes it a good fit for parallel resource access environments.

\subsection{Group Fairness Scheduling} \label{subsec:cluster_alloc}
We first establish the group resource allocation policy in a general multiuser framework.  The D2D setting will be covered subsequently as a specific case.  Consider a multiuser system where all the users share a single common resource.  There are a total of $K$ users in the system.  The set of users are partitioned into $G$ groups where group $i$ has $m_i$ users.  The groups compete for \emph{exclusive} access to the resource (orthogonal sharing).  All the users in each group has simultaneous access to the resource (non-orthogonal sharing) when the group is granted access.  Each user $j$ in group $i$ accesses the resource with the access metric $X_{i,j}$ (e.g., SNR, data rate, etc.).

There are two steps associated with a group allocation policy:
    \begin{itemize}
      \item Representative forming: the \emph{intra-group} process to form a \emph{representative} for each group.
      \item Representative selecting: the \emph{inter-group} selection process to grant resource to a single group based on the group representatives.
    \end{itemize}
In order to have a fair comparison between different users whose probability distributions can be very different, for each user $j$ in group $i$, the CDF-mapped value $V_{i,j} \triangleq F_{X_{i,j}}(X_{i,j})$ is used instead of the raw metric $X_{i,j}$ for all resource consideration purposes.  This is motivated by the use of the CDF value in the BCS policy. For simplicity, for each group $i$, we form the representative, $Y_i$, using the \emph{max representative} scheme as follows
    \begin{flalign}
    Y_i = \underset{j}{\op{max}} \, V_{i,j}. \label{eqn:max_rep}
    \end{flalign}
This scheme is clearly fair for all the group members as each has the same probability of being the group representative (recall that $V_{i,j}$'s are uniformly i.i.d. \cite{Park2005}).  Under this representative forming scheme, however, larger groups can have an advantage over smaller groups in terms of the representative $Y$.  As a result, it is necessary to weight each group differently in the inter-group selection step.  We employ the following \emph{Max Weighted Selection} (MWS) scheme
    \begin{flalign}
    i^{*} = \underset{i}{\op{argmax}} \, Y_{i}^{1/w_i}, \label{eqn:inter_cluster}
    \end{flalign}
where $w_i$ is selection weight for group $i$.  The main task in our scheduling problem is to choose the set of $w_i$'s to maximize the user performance while ensuring user fairness.  Following the UPI optimality property of CDF scheduling in theorem \ref{theorem:cdf_opt}, we adopt the max-min UPI optimization criteria to solve for the group weights
    \begin{flalign}
    \mathbf{w}^{*} = \underset{\mathbf{w}}{\op{argmax}} \, \underset{i,j}{\op{min}} \op{UPI}_{i,j}^{(X)} \text{ where } \mathbf{w} \triangleq [w_1, w_2, \dots, w_G]. \label{prob:max_min1}
    \end{flalign}
Problem (\ref{prob:max_min1}) can be reformulated as follows
    \begin{flalign*}
    \mathbf{w}^{*} = \underset{\mathbf{w}, u}{\op{argmin}} \, \left(\frac{1}{u}\right) \text{ s.t. } \op{UPI}_{i,j}^{(X)} \ge u, \forall i, j \Leftrightarrow \mathbf{w}^{*} = \underset{\mathbf{w}, u}{\op{argmin}} \, \left(\frac{1}{u}\right) \text{ s.t. } u \left[\op{UPI}_{i,j}^{(X)}\right]^{-1} \le 1, \forall i, j,
    \end{flalign*}
where $u > 0$ is a dummy optimization variable.  The individual UPIs and group access probabilities are given by theorems \ref{theorem:non_ortho_upi} and \ref{theorem:prob_non_ortho} below.
\begin{theorem}
\label{theorem:non_ortho_upi}
Under the MWS scheme, the UPI for user $j$ in group $i$ with size $m_i$ is given by
    \begin{flalign}
    \op{UPI}_{i,j}^{(X)} &= \frac{m_i + 1} {\mu_i + 1}
    \text{ where } \mu_i \triangleq \sum_{k=1}^G (m_k w_k)/w_i. \label{eqn:fi1}
    \end{flalign}
\end{theorem}

\begin{theorem}
\label{theorem:prob_non_ortho}
The probability of group $i$ being selected is given by 
    \begin{flalign}
    P_i = \frac{m_i}{\mu_i} = \frac{m_i w_i}{\sum_{k=1}^G m_k w_k}. \label{eqn:prob_non_ortho}
    \end{flalign}
\end{theorem}

%

Using (\ref{eqn:fi1}), we obtain the following optimization problem
    \begin{equation}
    \begin{gathered}
    \mathbf{w}^{*} = \underset{\mathbf{w}, u}{\op{argmin}} \, \left( \frac{1}{u} \right) \label{prob:max_min2}
    \text{ s.t. } \left(\frac{1}{m_i + 1}\right) u \left(\sum_{k=1}^G (m_k w_k)/w_i + 1 \right) \le 1, \forall i = 1, \dots, G.
    \end{gathered}
    \end{equation}

Problem (\ref{prob:max_min2}) is a standard \emph{Geometric Program}, which can be solved efficiently for the weight $w_i$.  The performance of users under this policy is stated by the following theorem.

\begin{theorem}
\label{theorem:cdf_non_ortho}
The CDF of the SNR for user $j$ in group $i$ when group $i$ is selected is given by
\begin{flalign}
  F_{S_{i,j}^{*}}(s) &= \frac{\mu_i(m_i - 1)}{m_i(\mu_i - 1)} F_{S_{i,j}}(s) + \frac{\mu_i - m_i}{m_i (\mu_i - 1)} \left[F_{S_{i,j}}(s)\right]^{\mu_i}, \, \forall s \ge 0. \label{eqn:cdf_non_ortho}
\end{flalign}
\end{theorem}


\begin{remark}
a) At the optimum of (\ref{prob:max_min2}), all users has the same UPI and thus the GFS policy is \emph{UPI fair}. b) The second term in (\ref{eqn:cdf_non_ortho}) shows the multiuser diversity gain.  When the group becomes large, $\mu_i$ becomes large, the multiuser diversity gain vanishes and the frequency reuse gain becomes dominant which manifests as a boost in the probability of access. c) Even though the Max Weighted Selection process (\ref{eqn:inter_cluster}) used for GFS resembles the user selection process in BCS, it is in fact very different: for BCS, the weights are predetermined based on the desired access probability while for GFS, the weights are computed by (\ref{prob:max_min2}) based on the system user dynamic to ensure overall fair access in a non-orthogonal sharing environment.
\end{remark}


\begin{table}[!ht]
    \vspace{-12pt}
    \centering
    \caption{GFS Policy Simulation Parameters}
    \begin{tabular}{ | l || c | c | c | c | c | c | c | c | c | c | c | c | c | c | }
    \hline
    Group & \multicolumn{1}{ c| }{1} & \multicolumn{7}{ c| }{2} & \multicolumn{2}{ c| }{3} & \multicolumn{4}{ c| }{4}  \\ \hline
    User & 1 & 2 & 3 & 4 & 5 & 6 & 7 & 8 & 9 & 10 & 11 & 12 & 13 & 14           \\ \hline
    Mean SNR & 100 & 60 & 70 & 5 & 16 & 40 & 20 & 2 & 4 & 40 & 36 & 80 & 7 & 40 \\ \hline
    Nakagami shaping param $m$ & 1 & 8 & 2 & 6 & 7 & 3 & 7 & 5 & 3 & 3 & 2 & 9 & 9 & 4 \\ \hline
    \end{tabular}
    \label{tab:rep_sched_sim_params}
    \vspace{-12pt}
\end{table}

\begin{figure}[!h]
\vspace{-10pt}
\centering
\begin{minipage}{.5\textwidth}
  \centering
  \includegraphics[width=1\textwidth, height=2in]{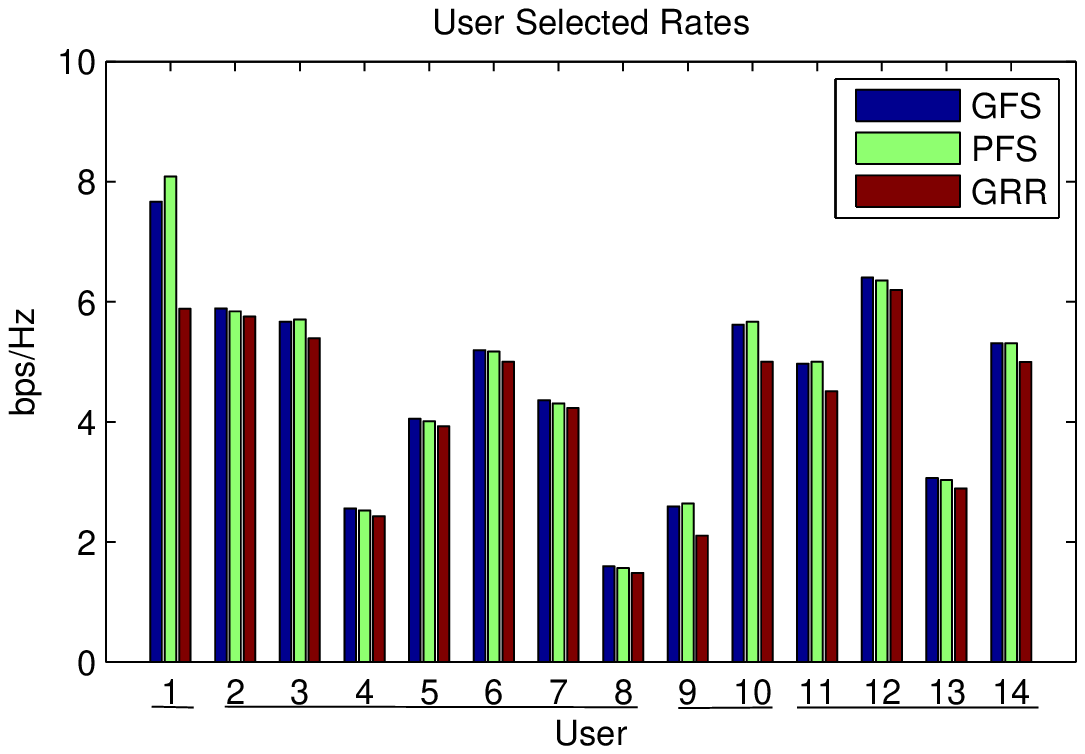}
  \captionof{figure}{User Selected Rates under GFS: all users outperform those under round-robin (GRR) policy}
  \label{fig:gfs_selected_rate}
\end{minipage}%
\begin{minipage}{.5\textwidth}
  \centering
  \includegraphics[width=1\textwidth, height=2in]{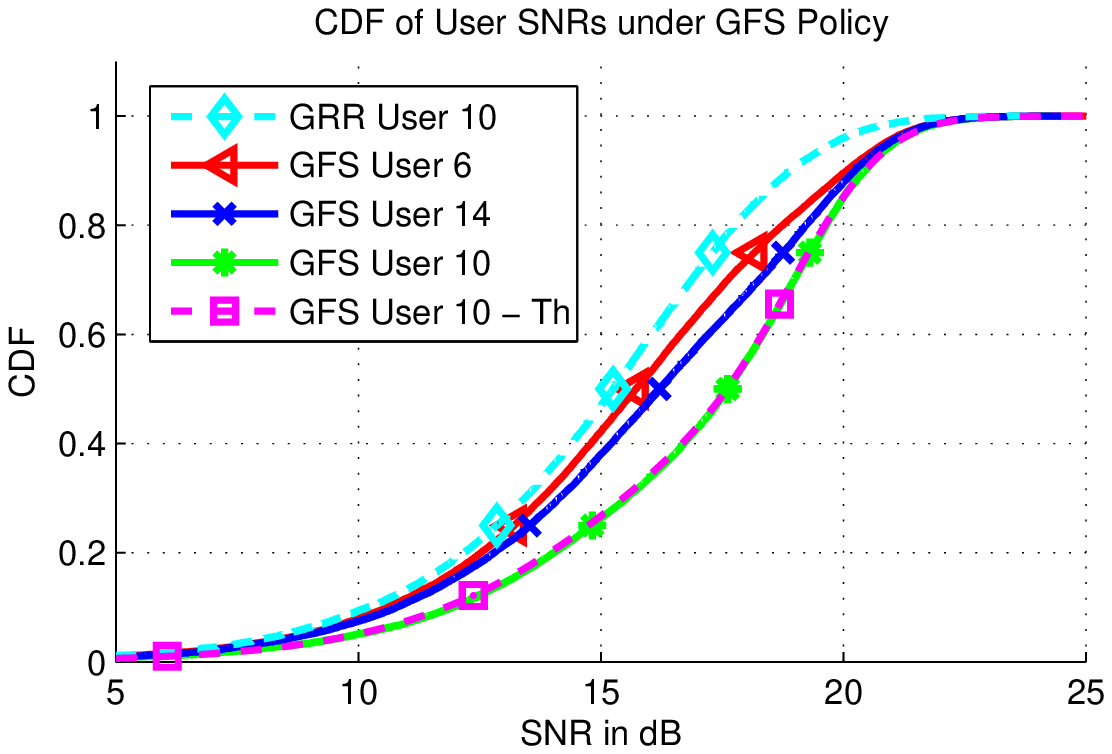}
  \captionof{figure}{User SNRs in Groups of Different Sizes under GFS: users in smaller groups get more diversity gain}
  \label{fig:gfs_snr_cdf}
\end{minipage}
\end{figure}

We now demonstrate the performance of the GFS policy in a small system.  For performance comparison purpose, we also consider a simple extension to the BCS policy called Extended CDF Scheduling (ECS), which maintains temporal fairness under group allocation.  This policy also uses the Max Weighted Selection scheme for group selection.  The temporal fairness can be achieved easily by setting $w_i = 1 / (m_i G)$, obtained by equating access probabilities given by (\ref{eqn:prob_non_ortho}).  The popular proportional fair (PFS) policy is also included.  To adapt for the group sharing setting, we use the max PF metric of each group as the group representative: $Y_i = \underset{j}{\op{max}} \, (X_{i,j} / \bar{X}_{i,j})$ where $\bar{X}_{i,j}$ is updated following the usual PF update rule, $\bar{X}_{i,j}(t) = (1 - 1/t_c) \bar{X}_{i,j}(t-1) + (1/t_c) X_{i,j}(t) \mathbf{1}_{i^{*}(t) = i}$.  When group $i$ is selected, the averages of all group members are updated with their current metrics.  Note that this adaptation is \emph{entirely heuristic} since the group members are updated whenever their group is granted even though they are not competing, i.e. their PF metrics are not the highest (except the group representative) as in the traditional PF scheme. Another simple policy, the Group Round-Robin Policy (GRR) where groups are granted resources in a round-robin fashion, is also included in the comparisons.  Assume the grouping of the users has been done yielding four user groups shown in table \ref{tab:rep_sched_sim_params}.  In order to simulate different user conditions, we assign each user an SNR value that follows a Nakagami-m distribution with a different (random) order $m$ and mean value.  In subsequent discussions, user rates are computed using Shannon formula: $\op{rate} = \op{log}(1 + \op{SNR})$.  The \emph{effective rate} of a user is the overall average rate the user is served, averaging across all times including the idle times when the user is not selected.  The \emph{selected rate} is the average of the instantaneous rates when the user is selected only.  The multiuser diversity gain under the GFS policy can be seen from figure \ref{fig:gfs_selected_rate} where all users outperform the round-robin policy (GRR).  Also seen from this figure, the GFS policy achieves similar multiuser diversity to the PFS policy. Figure \ref{fig:gfs_snr_cdf} shows the diversity gains for users in groups of different sizes.  As expected, users in larger groups get less diversity gain as indicated by theorem \ref{theorem:cdf_non_ortho}.  The theoretical CDF expression (\ref{eqn:cdf_non_ortho}) is also plotted on figure \ref{fig:gfs_snr_cdf}.



\begin{figure}[!t]
\vspace{-10pt}
\centering
\begin{minipage}{.5\textwidth}
  \centering
  \includegraphics[width=3.25in, height=2in]{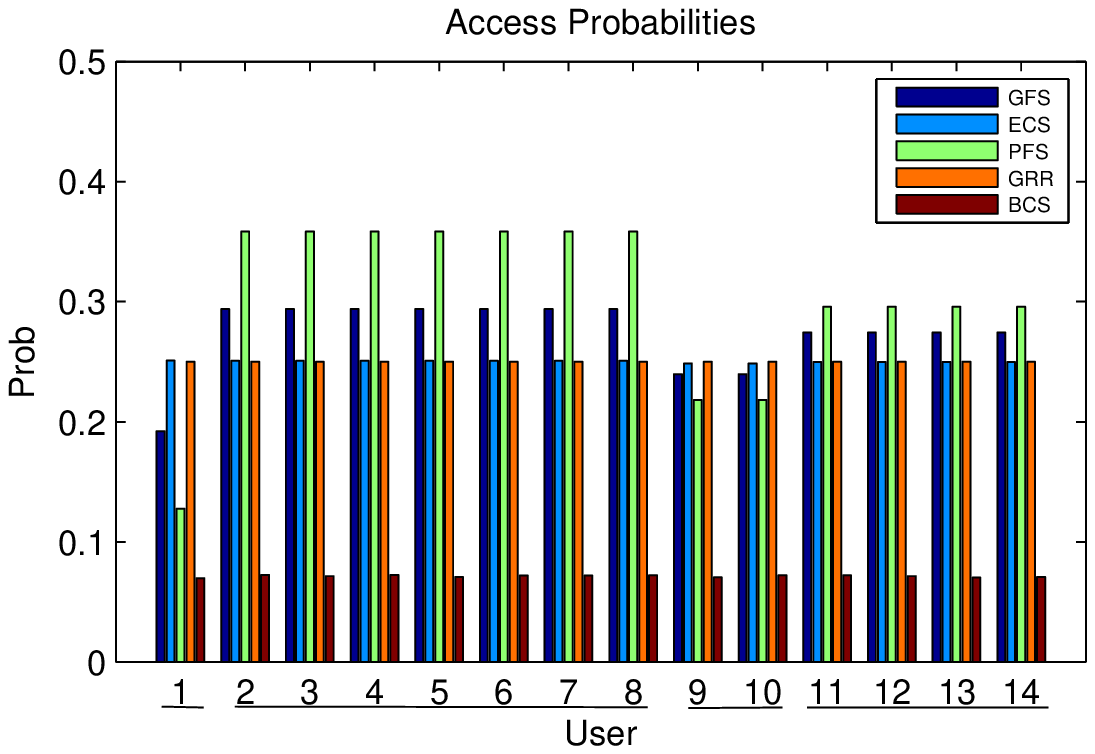}
  \captionof{figure}{Access Probability: GFS gives users in larger groups a boost while PFS overcompensates them}
  \label{fig:gfs_access_prob}
\end{minipage}%
\begin{minipage}{.5\textwidth}
  \centering
  \includegraphics[width=3.25in, height=2in]{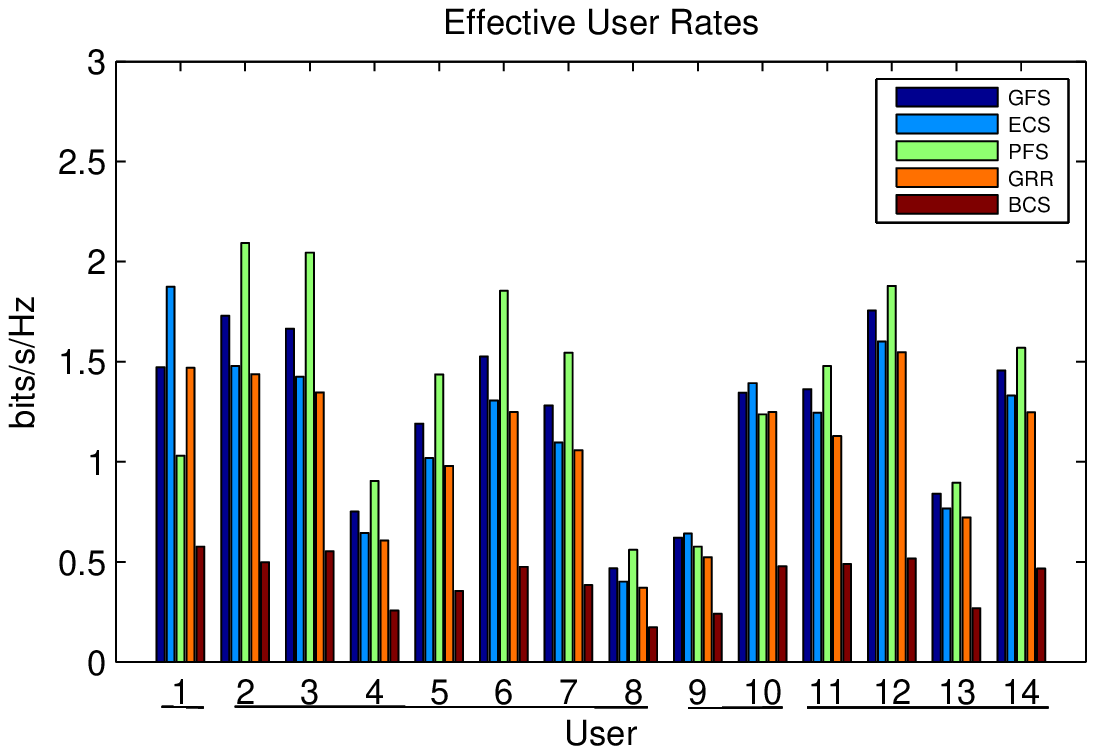}
  \captionof{figure}{User effective rates under GFS Policy: all users get better rates than round-robin}
  \label{fig:gfs_effective_rates}
\end{minipage}
\vspace{-15pt}
\end{figure}

\begin{figure}[!t]
\centering
\begin{minipage}{.5\textwidth}
  \centering
  \includegraphics[width=3.25in, height=2in]{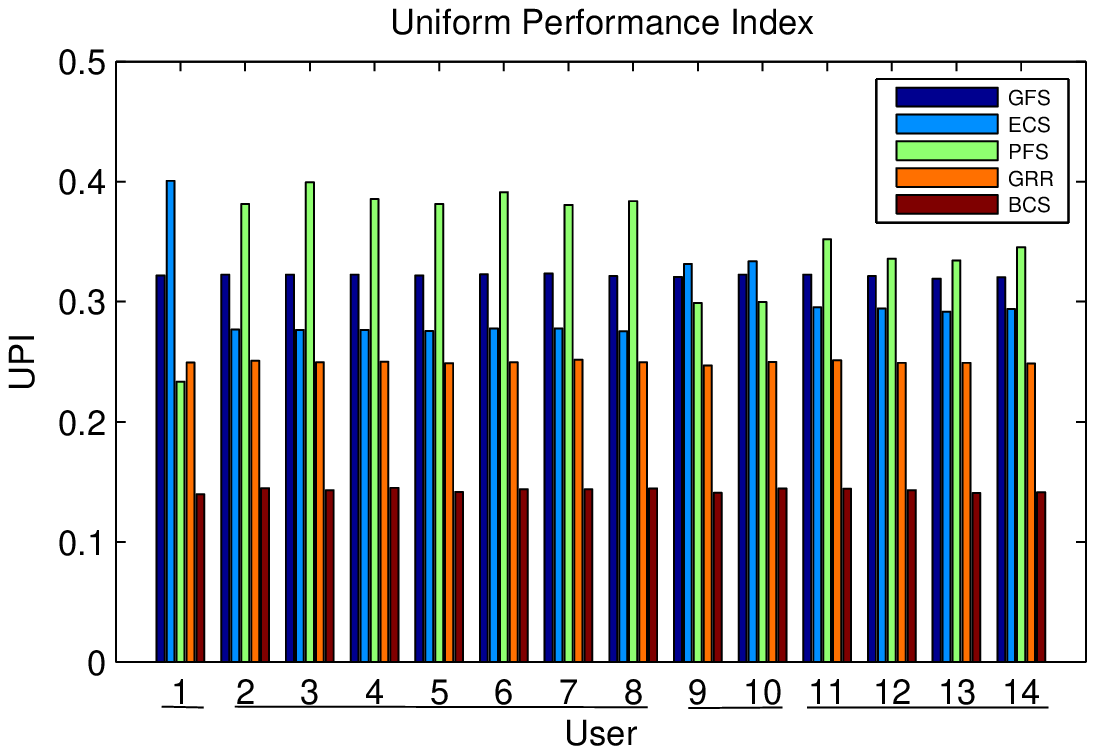}
  \captionof{figure}{UPI for different policies: the GFS policy is UPI-fair\\}
  \label{fig:gfs_upi}
\end{minipage}%
\begin{minipage}{.5\textwidth}
  \centering
  \includegraphics[width=3.25in, height=2in]{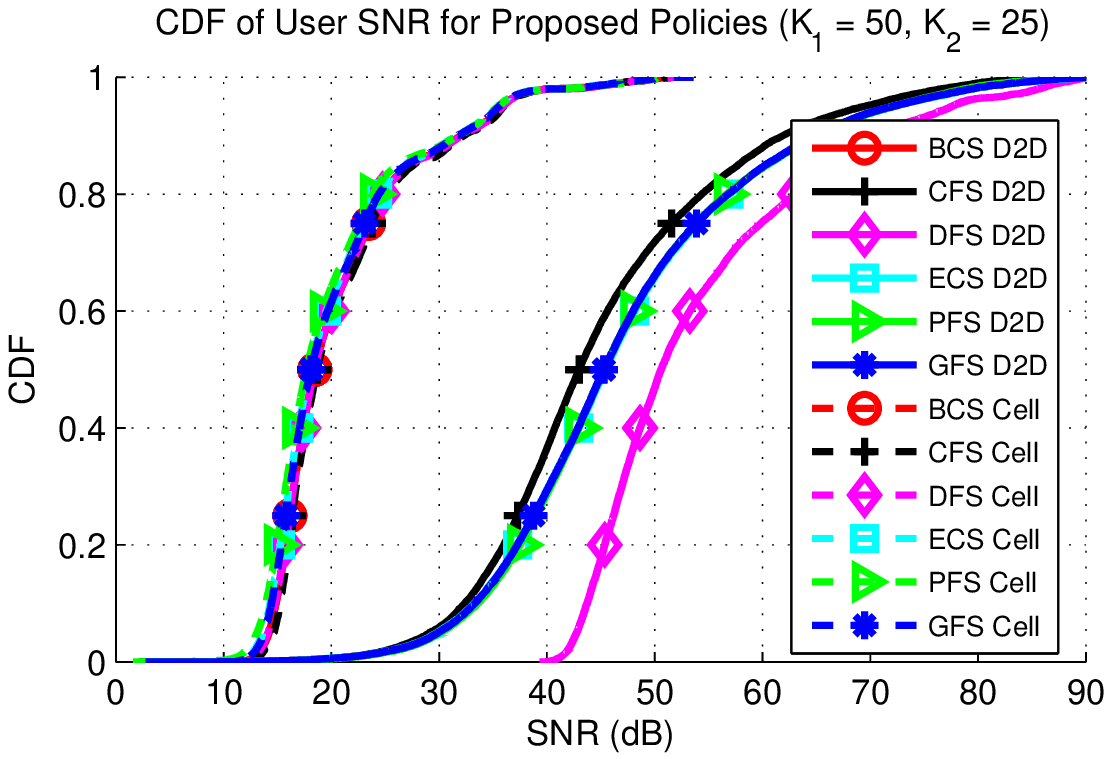}
  \captionof{figure}{D2D Diversity: DFS has the most gain, followed by GFS, then CFS}
  \label{fig:snr_cdf_all_m5}
\end{minipage}
\vspace{-20pt}
\end{figure}

In Figure \ref{fig:gfs_access_prob}, the effects of diversity loss compensation can be clearly seen.  Larger groups suffer more diversity loss and thus the GFS policy grants them a larger access probability.  The ECS policy, however, gives the same access time to all groups.  As expected, the orthogonal BCS scheme has much smaller user access probability due to orthogonal resource allocation.  The PFS policy, on the other hand, appears to "overcompensate" for the larger groups, leading to lower access time for small group users.  Figures \ref{fig:gfs_access_prob} and \ref{fig:gfs_effective_rates} also illustrate the non-orthogonal sharing gain achieved under the GFS policy.  The spectrum reuse results in large increments in user access probabilities, which translate to large improvements in user effective rates compared to an orthogonal scheme such as the BCS policy.  It is also clear that the GFS policy offers better performance/fairness tradeoff than the simple ECS policy. Under the ECS policy, users under small groups (users 1, 9, 10) have better performance at the cost of poorer performance for many users in large groups (such as users 2-8, 10-14).  This does not happen under the GFS policy: users under larger groups are properly compensated, leading to better connections for many more users.  Figure \ref{fig:gfs_effective_rates} also shows both diversity and non-orthogonal sharing gains achieved by the GFS policy as it outperforms both the group round-robin (GRR) and the orthogonal BCS policies.  It can also be seen from this figure that under the PFS policy, due to the overcompensation for larger groups, users in smaller groups do not get as much performance as those under the GFS policy while large group users enjoy a larger boost.  This is the main drawback of using the best group member metric as representative for group allocation discussed at the beginning of this section.  The UPI fairness of the GFS policy can be seen in figure \ref{fig:gfs_upi}.



\subsection{Group Fairness Scheduling For D2D} \label{subsec:gfs_for_d2d}
Let us now return to the resource allocation problem for D2D. When non-orthogonal sharing is allowed, the D2D pairs can be gathered in sharing groups and the results in subsection \ref{subsec:cluster_alloc} can be applied.  As in the DFS scheme in subsection \ref{subsec:d2d_fairness}, each D2D pair is considered as a single resource contender. Since the pair granted resource must be divided between the two users, the D2D UPI is only half of the value given in (\ref{eqn:fi1}) (assuming fair division):
\begin{flalign*}
    \op{UPI}_{i,j}^{(X)} &= \frac{1}{2} \times \frac{m_i + 1} {(\mu_i + 1)}
    \text{ where } \mu_i \triangleq \sum_{k} (m_k w_k)/w_i.
\end{flalign*}

As a result, problem (\ref{prob:max_min2}) becomes
\begin{equation}
    \begin{gathered}
    \mathbf{w}^{*} = \underset{\mathbf{w}, u}{\op{argmin}} \, \left( \frac{1}{u} \right) \label{prob:max_min3}
    \text{ s.t. } \left(\frac{1}{\nu_i(m_i + 1)}\right) u \left(\sum_{k} (m_k w_k)/w_i + 1 \right) \le 1.
    \end{gathered}
\end{equation}
where $\nu_i = 1$ for all cellular users and $\nu_i = 1/2$ for all D2D groups.  We propose the D2D Group Fairness Scheduling (GFS) policy in table \ref{tab:d2d_non_ortho}.

\begin{table}[!ht]
    \centering
    \caption{D2D Group Fairness Scheduling (GFS) Policy}
    \begin{tabular}{ | p{6.3in} |}
    \hline
    \begin{enumerate}
    \item Partition the D2D pairs into multiple sharing groups. \label{sched_d2d_partition}
    \item Solve (\ref{prob:max_min3}) to find the group weights. \label{d2d_non_ortho_weight_step}
    \item For each downlink/uplink frequency resource:
      \begin{enumerate}
        \item For each D2D group $i$, form $Y_i$ according to (\ref{eqn:max_rep}): $Y_i = \underset{j}{\op{max}} \, V_{i,j}$.
        \item Select a group according to (\ref{eqn:inter_cluster}): $i^{*} = \underset{i}{\op{argmax}} \, Y_{i}^{1/w_i}$.
      \end{enumerate}
    \item Grant access to all D2D pairs in a selected group.
    \item For each granted D2D pair, assign the resource to the two users according to the predefined ratio.
    \end{enumerate} \\ \hline
    \end{tabular}
    \label{tab:d2d_non_ortho}
    \vspace{-10pt}
\end{table}

Step \ref{sched_d2d_partition} in this policy involves forming D2D groups.  Given a number of D2D pairs distributed randomly in a geographic area, there are many ways the grouping can be done with different group memberships.  Different grouping results can lead to different system performances.  This grouping problem can be cast into a classical \emph{graph coloring problem}.  One can construct a \emph{conflict graph} where the vertices are the D2D pairs and the edges are drawn between any two pairs whose transmitters can cause significant interference to the other. The grouping problem then becomes one of coloring the vertices such that adjacent vertices have different colors.  Each color corresponds to a D2D sharing group.  Finding a coloring/grouping scheme with the smallest number of colors/groups is an NP-hard problem.  However, heuristic algorithms such as the \emph{greedy coloring algorithm} \cite{ChartrandAndZhang2010} can be used.  Whenever the group sizes are larger than one, non-orthogonal sharing gain can be realized.  With the low mobility of the D2D users, the first two steps of the policy are only done infrequently as the group structure does not change often.

\subsection{Simulation of All Three Proposed Scheduling Policies - A Performance Comparison}

\begin{figure}
\centering
\begin{minipage}{.5\textwidth}
  \centering
  \includegraphics[width=3.25in, height=2in]{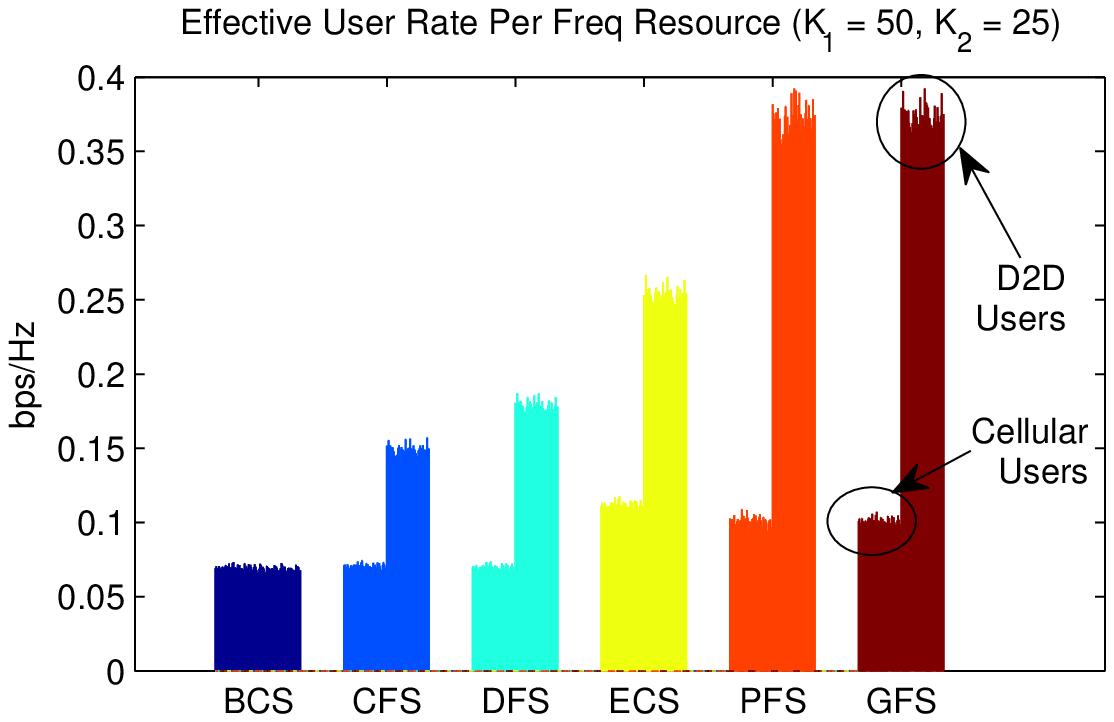}
  \captionof{figure}{Effective rates: GFS and PFS have highest performance, while DFS outperforms CFS, followed by BCS.}
  \label{fig:rs_vs_ortho}
\end{minipage}%
\begin{minipage}{.5\textwidth}
  \centering
  \includegraphics[width=3.25in, height=2in]{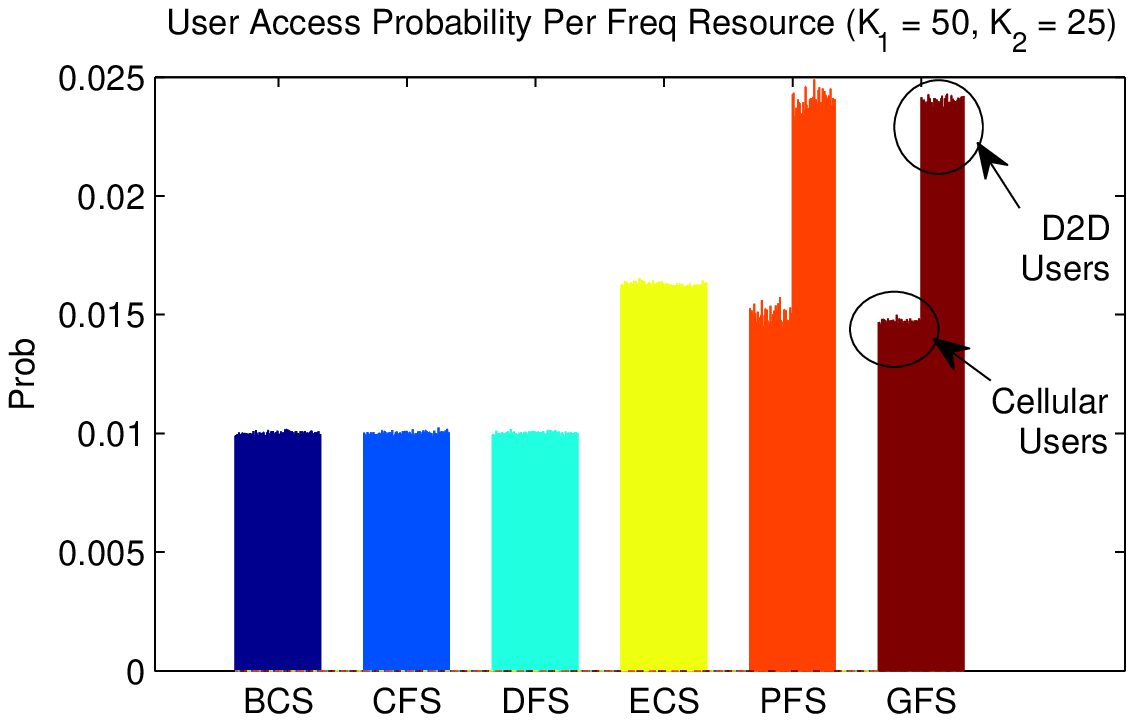}
  \captionof{figure}{Access probabilities: Users under GFS and PFS get a large boost in access time.}
  \label{fig:rs_vs_ortho_access}
\end{minipage}
\vspace{-20pt}
\end{figure}

In this section, we simulate a D2D system following the system model described in section \ref{sec:system_model}.  The simulation parameters are listed in table \ref{tab:sim_params} with $K_1 = 50$ and $K_2 = 25$.  The simulation is run for 150 realizations of the user spatial distribution and 12,000 realizations of fast fading per spatial realization.  The D2D users are grouped into 5 groups of 5 pairs each.  The reference ECS scheme is adjusted for D2D by setting $w_i = 2/(m_i G)$ to obtain access for both users.  The heuristic PFS scheme for the group setting described in subsection \ref{subsec:cluster_alloc} is also included.  Figure \ref{fig:snr_cdf_all_m5} shows the exploitation of multiuser diversity gains under different policies.  The orthogonal DFS policy gets the full D2D diversity gain.  The group policies GFS and ECS achieve less while the CFS policy receives no D2D diversity gain.  The PFS policy receives similar user diversity to the GFS policy.  The diversity gains for cellular users are the same under all policies.  Figure \ref{fig:rs_vs_ortho} shows the average user rates for all the proposed scheduling policies as well as the PFS policy.  As expected, the D2D users in both CFS and DFS policies receive much higher rates than they could under the BCS policy while the rates for cellular users remain unchanged.  The simplistic CFS policy performs very well compared to the DFS policy.  This is due to the fact that the D2D direct links typically provide very good throughput on average (D2D proximity gain).  The GFS policy, however, outperforms all other proposed policies and behaves similarly to the heuristic PFS policy.  Not only do the D2D users under this policy receive much higher throughput, but the cellular users also receive a big jump in their performance.  These gains can be attributed to the non-orthogonal sharing gain and the diversity loss compensation.  The non-orthogonal gain results in the increment in the probability of access for all users as evident from figure \ref{fig:rs_vs_ortho_access}.  The advantage of diversity loss compensation by group weighting in the GFS policy can be seen clearly in figure \ref{fig:rs_vs_ortho_access}.  Without this weighting, the D2D users would receive the same access probability as cellular users and thus would not receive as much rate gain.  It is worth noting that under this simulation setup with many cellular users and a relatively small group size of 5, the PFS policy has similar diversity loss compensation to the GFS policy (figure \ref{fig:rs_vs_ortho_access}), resulting in the performance similarity (figure \ref{fig:rs_vs_ortho}).  When the D2D groups have different sizes, the PFS scheme tends to overcompensate for users in larger groups as seen in figure \ref{fig:gfs_access_prob}, leading to unfair gains for D2D users in larger groups. On the contrary, the GFS policy performs fairer compensation based on UPI fairness.  In addition, under the GFS policy, user fairness can be adjusted easily by changing the optimization weights ($\nu_i$) in (\ref{prob:max_min3}).  For the PFS policy, however, adjusting user fairness or priority is much more difficult.  This typically requires a manual, imprecise process of online parameter tuning, which results in the use of approximate values leading to loss in performance.




\section{Conclusion} \label{sec:conclusion}
In this paper we first analyze and demonstrate D2D performance gains under two simple orthogonal scheduling policies.  For non-orthogonal environments, we introduce the UPI concept and propose the \emph{group fairness scheduling} (GFS) policy.  As evident from our analysis and simulations, the GFS policy provides not only perfect fairness but also excellent rate performance for all users.  Many salient features of this policy make it very well suited for D2D environments.



%

\appendices

\section{Proof of Theorem \ref{theorem:cdf_opt}} \label{app:theorem:cdf_opt}
We need the following lemmas to prove the result in theorem \ref{theorem:cdf_opt}:
\begin{lemma}
\label{lemma:bcs_max_supi}
The BCS policy maximizes the system total UPI (sUPI).  That is,
    \begin{flalign*}
      \op{BCS} &= \underset{\pi \in \mathcal{P}}{\operatorname{argmax}} \, \op{sUPI}^{\pi},
    \end{flalign*}
    where $\mathcal{P}$ is the set of all scheduling policies.
\end{lemma}
{\setlength{\parindent}{0in}
\emph{Proof.} Under the BCS scheme, at time instance $n$, the user with maximum $V_i[n] = F_{X_i}(X_i[n])$ is selected.  Let  $i^{*}[n] = \underset{i}{\operatorname{argmax}} \, V_i[n]$ and $V^{*}[n] = V_{i^{*}}[n]$.  Since only one of the $K$ users is selected $\sum_{i=1}^{K} U_i^{*}[n] = V^{*}[n]$ where $U_i^{*}[n] = V_i[n] 1_{i^{*} = i}$.  Assuming stationarity, we have
}
    \begin{flalign*}
    E[V^{*}] &= E[V^{*}[n]] = E\left[ \sum_{i = 1}^{K} U_i^*[n] \right]
    = \sum_{i = 1}^{K} E\left[ U_i^*[n] \right] = \sum_{i = 1}^{K} \frac{1}{2} \op{UPI}_i^{(X)} = \frac{1}{2} \op{sUPI}^{\op{BCS}},
    \end{flalign*}
    where $V^{*}$ is the system selected CDF-mapped value.  Assuming ergodicity, we have
    \begin{flalign*}
    \op{sUPI}^{\op{BCS}} = 2 E[V^{*}] &= 2 \underset{N \rightarrow \infty}{\op{lim}} \, \frac{1}{N} \sum_{n = 1}^{N} V^{*}[n].
    \end{flalign*}

Let $W^{*}[n] = F_{X_{j^{*}}}(X_{j^{*}}[n])$ be the CDF-mapped value of any other scheduling policy $\pi$, where $X_{j^{*}}[n]$ is the value of the selected user.  Since $V^{*}[n]$ is the maximum according to the BCS policy:
    \begin{flalign*}
    W^{*}[n] &\le V^{*}[n] \Rightarrow \frac{1}{N} \sum_{n = 1}^{N} W^{*}[n]  \le \frac{1}{N} \sum_{n = 1}^{N} V^{*}[n]\\
    E[W^{*}] &= \underset{N \rightarrow \infty}{\op{lim}} \, \frac{1}{N} \sum_{n = 1}^{N} W^{*}[n] \le \underset{N \rightarrow \infty}{\op{lim}} \, \frac{1}{N} \sum_{n = 1}^{N} V^{*}[n] = E[V^{*}]     \Rightarrow \op{sUPI}^{\pi} \le \op{sUPI}^{\op{BCS}}. \qed
    \end{flalign*}

\begin{lemma}
\label{lemma:bcs_equal_upi}
All users receive the same UPI of $\frac{2}{K + 1}$ under the BCS policy.
\end{lemma}
{\setlength{\parindent}{0in}
\emph{Proof.} The CDF of $U_i^{*}$ from definition \ref{def:upi} is given by
    \begin{flalign*}
    &F_{U_i^{*}}(u) = \op{Pr}[U_i^{*} < u; \text{i is selected}] + \op{Pr}[U_i^{*} < u; \text{i is not selected}] \\
    &= \op{Pr}[U_i^{*} < u; \text{i is selected}] + \underbrace{\op{Pr}[U_i^{*} < u| \text{i is not selected}]}_{1} \underbrace{\op{Pr}[\text{i is not selected}]}_{C} \\
    &= \op{Pr}[U_i < u; U_j < U_i, \forall j \ne i] + C \overset{(a)}{=} \int_{0}^{1} \op{Pr}[t < u; U_j < t, \forall j \ne i| U_i = t] f_{U_i}(t) dt + C\\
    &= \int_{0}^{u} \op{Pr}[U_j < t, \forall j \ne i] f_{U_i}(t) dt  + C \overset{(b)}{=} \int_{0}^{u} t^{K-1} dt = \frac{u^K}{K} + C \Rightarrow f_{U_i^{*}}(u) = u^{K-1} + C \delta(u) \\
    &\op{UPI}_i^{(X)} = 2\mathbf{E} \{ U_i^{*} \} = 2\int_{0}^{1} u f_{U_i^{*}}(u) du = 2 \int_{0}^{1} u u^{K-1} du = \frac{2}{K + 1}, \forall i. \qed
    \end{flalign*}
    where (a) is from the law of total probability and (b) is from the independence of $U_j$ and the properties of CDF/PDF of uniform random variables.  Lemmas \ref{lemma:bcs_max_supi} and \ref{lemma:bcs_equal_upi} lead directly to the max-min optimality result in theorem \ref{theorem:cdf_opt}.

\section{Proof of Theorem \ref{theorem:cdf_snr}} \label{app:theorem:cdf_snr}
The CDF of the user SNR $S_{k,c}$ when it is selected is given by
\begin{flalign*}
F_{S_{k,c}}^{*}(s) &= \op{Pr}[S_{k,c} < s| U_k > U_j, \forall j \ne k] = \op{Pr}[S_{k,c} < s; U_k > U_j, \forall j \ne k] / \op{Pr}[U_k > U_j, \forall j \ne k] \\
&\overset{(a)}{=} \op{Pr}[F_{S_{k,c}}(S_{k,c}) < F_{S_{k,c}}(s); U_k > U_j, \forall j \ne k] / (1/K) \\
&= K \op{Pr}[U_k < F_{S_{k,c}}(s); U_k > U_j, \forall j \ne k] \overset{(b)}{=} K \int_0^{F_{S_{k,c}}(s)} \left(\prod_{\forall j \ne k} \op{Pr}[U_j < u]\right) f_{U_k}(u) du \\
&= K \int_0^{F_{S_{k,c}}(s)} \left(\prod_{\forall j \ne k} F_{U_j}(u)\right) f_{U_k}(u) du \overset{(c)}{=} \int_0^{F_{S_{k,c}}(s)} K u^{K-1} du = [F_{S_{k,c}}(s)]^K. \qed
\end{flalign*}
where (a) is from the fact that each user is selected with probability $1/K$; (b) from the independence of $U_j$'s; (c) from the fact that $U_k$ and $U_j$'s are uniformly distributed in $[0, 1]$.

\section{Proof of Theorem \ref{theorem:cell_fairness_prob}} \label{app:theorem:cell_fairness_prob}
With $U_k = F_{S_{k,c}}(S_{k,c})$, the probability of selecting a cellular user is
\begin{flalign*}
P_c &= \op{Pr}[U_k \ge u^{th}, \text{for some } k \in \mathcal{K}_c] = 1 - \op{Pr}[U_k < u^{th}, \forall k \in \mathcal{K}_c]\\
&= 1 - \prod_{k \in \mathcal{K}_c} \op{Pr}[U_k < u^{th}] = 1 - \prod_{k \in \mathcal{K}_c} F_{U_k}(u^{th}) \overset{(a)}{=} 1 - \left(u^{th}\right)^{K_1} = 1 - \left[(K-K_1)/K\right] = \frac{K_1}{K},
\end{flalign*}
where (a) is from the CDF of uniform random variables.  Thus, each of the $K_1$ cellular users has access probability of $P_c / K_1 = 1/K$.  This is the same for each D2D user.

\section{Proof of Theorem \ref{theorem:cell_fairness_opt}} \label{app:theorem:cell_fairness_opt}

Let $\Omega$ be the set of all scheduling policies that are temporally fair to cellular users.  Consider an arbitrary policy $\pi \in \Omega$.  Let $W$ be a time window where $W$ scheduling decisions are made.  Within this window, let $N = N_1 + N_2$ be the number of cellular user selections, where $N_2$ selections are made when all cellular statistics are below the threshold $u_k < u^{th}$.  Let $\mathcal{N}$, $\mathcal{N}_1$, $\mathcal{N}_2$ be the corresponding sets of time indices, where $\mathcal{N} = \mathcal{N}_1 \cup \mathcal{N}_2$.  During this same window, consider the CFS policy.  Let $M = M_1 + M_2$ be the number of cellular user selections under CFS.  Likewise, let $\mathcal{M}$, $\mathcal{M}_1$, $\mathcal{M}_2$ be the corresponding sets of time indices, where $\mathcal{M} = \mathcal{M}_1 \cup \mathcal{M}_2$.  Let $\mathcal{M}_1 = \mathcal{N}_1$ (and thus $M_1 = N_1$), which is possible since CFS always selects a cellular user when at least one $u_k \ge u^{th}$.  We consider the following two cases:  $M_2 \ge N_2$ and $M_2 < N_2$.  Let $u^{*}[n]$ be the statistic for the selected cellular user under $\pi$ at time $n$.  In the first case when $M_2 \ge N_2$, we always have
\begin{flalign*}
\frac{1}{W} \sum_{n \in \mathcal{N}} u^{*}[n] &= \frac{1}{W} \sum_{n \in \mathcal{N}_1} u^{*}[n] + \frac{1}{W} \sum_{n \in \mathcal{N}_2} u^{*}[n] \le \frac{1}{W} \sum_{n \in \mathcal{N}_1} \op{max} u_k [n] + \frac{1}{W} \sum_{n \in \mathcal{N}_2} u^{th} \\
&\le \frac{1}{W} \sum_{n \in \mathcal{M}_1} \op{max} u_k [n] + \frac{1}{W} \sum_{n \in \mathcal{M}_2} \op{max} u_k [n] = \frac{1}{W} \sum_{n \in \mathcal{M}} \op{max} u_k [n]\\
\Rightarrow \op{cUPI}^{\pi} &= 2 \underset{W \rightarrow \infty}{\op{lim}} \frac{1}{W} \sum_{n \in \mathcal{N}} u^{*}[n] \le 2 \underset{W \rightarrow \infty}{\op{lim}} \frac{1}{W} \sum_{n \in \mathcal{M}} \op{max} u_k [n] = \op{cUPI}^{CFS}, \numberthis \label{eqn:cUPI1}
\end{flalign*}
where $\op{cUPI}$ is the total UPI of all cellular users.  In the second case when $M_2 < N_2$, let $N_2 = M_2 + N_3$, and divide into $\mathcal{N}_2 = \hat{\mathcal{N}}_2 \cup \mathcal{N}_3$ where $|\hat{\mathcal{N}}_2| = M_2$ and $|\mathcal{N}_3| = N_3$, we have
\begin{flalign*}
&\frac{1}{W} \sum_{n \in \mathcal{N}} u^{*}[n] = \frac{1}{W} \sum_{n \in \mathcal{N}_1} u^{*}[n] + \frac{1}{W} \sum_{n \in \hat{\mathcal{N}}_2} u^{*}[n] + \frac{1}{W} \sum_{n \in \mathcal{N}_3} u^{*}[n] \\
&\le \frac{1}{W} \sum_{n \in \mathcal{N}_1} \op{max} u_k [n] + \frac{1}{W} \sum_{n \in \hat{\mathcal{N}}_2} u^{th} + \frac{1}{W} \sum_{n \in \mathcal{N}_3} u^{th}
\le \frac{1}{W} \sum_{n \in \mathcal{M}} \op{max} u_k [n] + \frac{N_3 u^{th}}{W}\\
\Rightarrow \op{cUPI}^{\pi} &= 2 \underset{W \rightarrow \infty}{\op{lim}} \frac{1}{W} \sum_{n \in \mathcal{N}} u^{*}[n] \le 2\underset{W \rightarrow \infty}{\op{lim}} \frac{1}{W} \sum_{n \in \mathcal{M}} \op{max} u_k [n] + 2\underset{W \rightarrow \infty}{\op{lim}} \frac{N_3 u^{th}}{W} \overset{(a)}{=} \op{cUPI}^{CFS}. \numberthis \label{eqn:cUPI2}
\end{flalign*}

Here, (a) results from the temporal fairness of both $\pi$ and CFS:
\begin{flalign*}
\underset{W \rightarrow \infty}{\op{lim}} \frac{N}{W} &= \underset{W \rightarrow \infty}{\op{lim}} \frac{M}{W} = \frac{K_1}{K} \Rightarrow \underset{W \rightarrow \infty}{\op{lim}} \frac{N-M}{W} = 0 \Rightarrow \underset{W \rightarrow \infty}{\op{lim}} \frac{N_3}{W} = 0.
\end{flalign*}

From (\ref{eqn:cUPI1}) and (\ref{eqn:cUPI2}), we conclude that the CFS policy yields the largest total UPI for all cellular users.  In addition, the following lemma \ref{lemma:cfs_equal_upi} states that under CFS, all cellular users receive the same UPI.  Consequently, the CFS policy is max-min optimal with respect to the UPI metric.

\begin{lemma}
\label{lemma:cfs_equal_upi}
All cellular users receive the same UPI of $\frac{2\left[1 - (u^{th})^{(K_1 + 1)}\right]}{K_1 + 1}$ under the CFS policy.
\end{lemma}
{\setlength{\parindent}{0in}
\emph{Proof.} The proof can be obtained following a procedure similar to the proof of lemma \ref{lemma:bcs_equal_upi}.
}


\section{Proof of Theorem \ref{theorem:cell_fairness_cdf}} \label{app:theorem:cell_fairness_cdf}
Let $\mathcal{E}_{k}$ be the event user $k$ is selected.  From Theorem \ref{theorem:cell_fairness_prob}, $\op{Pr}[\mathcal{E}_k] = 1/K$.  With $u^{th} = \left(\frac{2K_2}{K}\right)^{1/K_1}$ and $S_{k,c}$ being the SNR for cellular user $k \in \mathcal{K}_c$ and $U_k = F_{S_{k,c}}(S_{k,c})$, we have
  \begin{flalign*}
  &\op{Pr}[S_{k,c} < s; \mathcal{E}_k] = \op{Pr}[S_{k,c} < s; U_k > U_j, \forall j \in \mathcal{K}_c, j \ne k; U_k > u^{th}]\\
  &= \op{Pr}[U_k < F_{S_{k,c}}(s); U_k > U_j, \forall j \in \mathcal{K}_c, j \ne k; U_k > u^{th}] \\
  &= \mathbf{1}_{(F_{S_{k,c}}(s) \ge u^{th})} \int_{u^{th}}^{F_{S_{k,c}}(s)} \op{Pr}[U_j < u, \forall j \in \mathcal{K}_c, j \ne k] f_{U_k}(u) du \\
  &\overset{(a)}{=} \mathbf{1}_{(F_{S_{k,c}}(s) \ge u^{th})} \int_{u^{th}}^{F_{S_{k,c}}(s)} u^{K_1 - 1} du
  = \mathbf{1}_{(F_{S_{k,c}}(s) \ge u^{th})} \times \frac{1}{K_1} \left[\left(F_{S_{k,c}}(s)\right)^{K_1} - (u^{th})^{K_1}\right] \\
  &= \frac{1}{K_1} \left[\left(F_{S_{k,c}}(s)\right)^{K_1} - \frac{2K_2}{K}\right]^{+}
  = \frac{1}{K} \left[\frac{K}{K_1}\left(F_{S_{k,c}}(s)\right)^{K_1} - \frac{2K_2}{K_1}\right]^{+}\\
  \Rightarrow F_{S_{k,c}^{*}}(s) &= \op{Pr}[S_k < s | \mathcal{E}_k] \overset{(b)}{=} \op{Pr}[S_k < s; \mathcal{E}_k] / \op{Pr}[\mathcal{E}_k]
  = \left[\frac{K}{K_1}\left(F_{S_{k,c}}(s)\right)^{K_1} - \frac{2K_2}{K_1}\right]^{+}, \qed
  \end{flalign*}
where (a) is from independent uniform variables, (b) from Bayes' rule.  For D2D users, $k \in \mathcal{K}_d$:
  \begin{flalign*}
  &\op{Pr}[S_{k,d} < s; \mathcal{E}_k] = \op{Pr}[S_{k,d} < s; U_j \le u^{th}, \forall j \in \mathcal{K}_c; k \in \mathcal{K}_d \text{ selected } ]\\
  &= \op{Pr}[S_{k,d} < s] \op{Pr}[U_j \le u^{th}, \forall j \in \mathcal{K}_c] \op{Pr}[k \in \mathcal{K}_d \text{ selected } ] = F_{S_{k,d}}(s) \left(\frac{2K_2}{K}\right) \left(\frac{1}{2K_2}\right)\\
  &= \frac{1}{K} F_{S_{k,d}}(s) \Rightarrow F_{S_{k,d}^{*}}(s) = \op{Pr}[S_k < s | \mathcal{E}_k] = F_{S_{k,d}}(s). \qed
  \end{flalign*}

\section{Proof of Theorem \ref{theorem:d2d_fairness_cdf}} \label{app:theorem:d2d_fairness_cdf}

Let $\mathcal{E}_{k}$ be the event user $k$ is selected.  When a cellular user is selected, we have
    \begin{flalign*}
    &Pr[S_{k^{*}} < s; \mathcal{E}_{k}] = Pr[S_{k^{*}} < s; k^{*} = k, k \in \mathcal{K}_c]
    \overset{(a)}{=} Pr[S_{k,c} < s; U_i^{1/w_i} < U_k^{1/w_k}, \forall i \ne k] \\
    &\overset{(b)}{=} Pr[F_{k,c}(S_{k,c}) < F_{k,c}(s); U_i^{1/w_i} < U_k^{1/w_k}, i \ne k]
    \overset{(c)}{=} Pr[U_{k} < F_{k,c}(s); U_i < U_k^{w_i/w_k}, i \ne k] \\
    &\overset{(d)}{=} \int_{0}^{1} Pr[u < F_{k,c}(s); U_{i} < u^{\frac{w_i}{w_k}}, i \ne k | U_k = u]f_{U_{k}}(u) du
    \overset{(e)}{=} \int_{0}^{F_{k,c}(s)} \prod_{i \ne k} Pr[U_{i} < u^{\frac{w_i}{w_k}}] du \\
    &\overset{(f)}{=} \int_{0}^{F_{k,c}(s)} u^{\sum_{i \ne k}\frac{w_i}{w_k} } du \overset{(g)}{=} \int_{0}^{F_{k,c}(s)} u^{\frac{1}{w_k} - 1} du
    = w_k \left[F_{k,c}(s) \right]^{1/w_k} = \frac{1}{K}\left[F_{k,c}(s) \right]^{K}\\
    &\Rightarrow F_{S_{k,c}^{*}}(s) = Pr[S_{k^{*}} < s | \mathcal{E}_{k}] = Pr[S_{k^{*}} < s; \mathcal{E}_{k}] / Pr[\mathcal{E}_{k}] = \left[F_{k,c}(s) \right]^{K}. \qed
    \end{flalign*}
Here we have (a) from the CDF-based selection criterion and the fact that user $k$ is selected; (b) from the monotonicity of CDF function; (c) from definition of random variables $U_{k}$; (d) from the law of total probability; (e) from independence of $U_{i}$; (f) from the definition of CDF, (g) from the fact $\sum_{i} w_i = 1$. Similarly, when a D2D pair is selected, with $Pr[\mathcal{E}_{k}] = w_k = 2/K$:
    \begin{gather*}
    Pr[S_{k^{*}} < s; \mathcal{E}_{k}] = Pr[S_{k,d} < s;  U_i^{1/w_i} < U_k^{1/w_k}, \forall i \ne k]
    = w_k \left[F_{k,d}(s) \right]^{1/w_k} \\
    \Rightarrow F_{S_{k,d}^{*}}(s) = Pr[S_{k^{*}} < s| \mathcal{E}_{k}] = \left[F_{k,d}(s) \right]^{1/w_k} = \left[F_{k,d}(s) \right]^{K/2}. \qed
    \end{gather*}

\section{Proof of Theorem \ref{theorem:dfs_uncond_cdf}} \label{app:theorem:dfs_uncond_cdf}
%

From (\ref{eqn:uncond_cdfa}), using binomial expansion, we have
  \begin{flalign*}
    &F_{S_{i,c}^{*}}(s) = \int \left[1 - \operatorname{exp} \left(-A_c s y^{\eta_c} \right) \right]^{K}  f_{Y}(y) dy
    = \sum_{i=0}^{K} {K\choose i} (-1)^{i} \int \op{exp} \left(-i A_c s y^{\eta_c} \right) f_{Y}(y) dy \\
    &= 1 + \sum_{i=1}^{K} {K\choose i} (-1)^{i} \int \op{exp} \left(-i A_c s y^{\eta_c} \right) f_{Y}(y) dy.
  \end{flalign*}

Using the user distributions in subsection \ref{ssubsec:usr_model}, $f_Y(y) = \frac{2 y}{R_B^2}$, we have
  \begin{flalign*}
    F_{S_{i,c}^{*}}(s) &= 1+ \sum_{i=1}^{K} {K\choose i} (-1)^{i} \frac{2}{R_B^2} \underbrace{\int_{0}^{R_B} \op{exp} \left(-i A_c s y^{\eta_c} \right) y dy}_{I_y}.
  \end{flalign*}

Using the change of variable $t = i A_c s y^{\eta_c}$, after some manipulations, we get
\begin{flalign*}
    I_y &= \frac{1}{\eta_c} \left( i A_c s \right)^{-2/\eta_c} \gamma(2/\eta_c, \beta_c(i)) \\
    F_{S_{i,c}^{*}}(s) &= 1 + \sum_{i=1}^{K} {K\choose i} (-1)^{i} \frac{2}{\eta_c R_B^2 \left( i A_c s \right)^{2/\eta_c}} \gamma(2/\eta_c, \beta_c(i)), \qed
  \end{flalign*}
  where $\gamma(a, x) = \int_{0}^{x} t^{a - 1} e^{-t} dt$, the lower incomplete gamma function.  For D2D users, from (\ref{eqn:uncond_cdfb})
    \begin{flalign*}
    F_{S_{i,d}^{*}}(s) &= 1 + \sum_{i=1}^{\infty} {\frac{K}{2} \choose i} (-1)^{i} \int \operatorname{exp} \left(-i A_d s x^{\eta_d} \right) f_{X}(x) dx.
    \end{flalign*}

Using $f_X(x) = \frac{1}{\Delta_D}$ and the change of variable $t = A_d s x^{\eta_d}$, after some manipulations, we get
    \begin{flalign*}
    &F_{S_{i,d}^{*}}(s) = 1 + \sum_{i=1}^{\infty} {K/2 \choose i} (-1)^{i} \frac{1}{\eta_d \Delta_D \left( i A_d s \right)^{1/\eta_d}} \left[\gamma(1/\eta_d, \beta_d(i)) - \gamma(1/\eta_d, \alpha_d(i)) \right], \qed
    \end{flalign*}
    where $\alpha_{d,i} = i A_d s D_{min}^{\eta_d}$, $\beta_{d,i} = i A_d s D_{th}^{\eta_d}$.

\section{Proof of Theorem \ref{theorem:non_ortho_upi}} \label{app:theorem:non_ortho_upi}
Letting $U_{i,j} = F_{V_{i,j}}(V_{i,j})$ for user $j$ in group $i$ and $U_{i,j}^{*} = U_{i,j} \mathbf{1}_{\{]\text{group i selected}\}}$, we have
    \begin{flalign*}
    &F_{U_{i,j}^{*}}(u) = \op{Pr}[U_{i,j}^{*} < u] = F(u) + \op{Pr} [\text{group i is not selected}] = F(u) + C W(u)\\
    &F(u) \triangleq \op{Pr}[U_{i,j}^* < u\, \op{and}\, \text{(group i is selected)}] = \op{Pr}[U_{i,j} < u; (Y_k)^{1/w_k} < (Y_i)^{1/w_i}, \forall k \ne i] \\
    &= \underbrace{\op{Pr}[U_{i,j} < u; (Y_k)^{1/w_k} < (Y_i)^{1/w_i}, \forall k \ne i; Y_i = V_{i,j}]}_{F_1(u)} \text{ ($V_{i,j}$ is the group representative)}\notag\\
    &+ \underbrace{\op{Pr}[U_{i,j} < u; (Y_k)^{1/w_k} < (Y_i)^{1/w_i}, \forall k \ne i; Y_i \ne V_{i,j}]}_{F_2(u)} \text{ ($V_{i,j}$ is not the group representative)},
    \end{flalign*}
    where $C$ is a constant and $W(u) = \mathbf{1}_{\{u \in [0, 1]\}}$.  Since $V_{i,j}$ is uniform, we have $U_{i,j} = V_{i,j}$ and
    \begin{flalign*}
    &F_1(u) = \op{Pr}[V_{i,j} < u; (Y_k)^{1/w_k} < (V_{i,j})^{1/w_i}, \forall k \ne i; Y_i = V_{i,j}] \\
    &= \op{Pr}[V_{i,j} < u; (Y_k)^{1/w_k} < (V_{i,j})^{1/w_i}, \forall k \ne i; V_{i,l} < V_{i,j}, l \ne j] \\
    &\overset{(a)}{=} \int_{0}^{1} \op{Pr}[v < u; (Y_k)^{1/w_k} < (v)^{1/w_i}, \forall k \ne i; V_{i,l} < v, \forall l \ne j | V_{i,j} = v] f_{V_{i,j}}(v) dv \\
    &\overset{(b)}{=} \int_{0}^{u} (v)^{\sum_{k} (m_k w_k)/w_i - 1} dv = \int_{0}^{u} (v)^{\mu_i - 1} dv, \text{ where }\mu_i \triangleq \sum_{k=1}^G (m_k w_k)/w_i. \text{ Similarly,}\\
    &F_2(u) = \op{Pr}[V_{i,j} < u; (Y_k)^{1/w_k} < (Y_{i^{*}})^{1/w_i}, \forall k \ne i; V_{i,j} < Y_{i^{*}}], \text{ where $Y_{i^{*}}$ is the representative.}\\
    &\text{Since }Y_{i^{*}} = \underset{l \ne j}{\op{max}} \, V_{i,l} \Rightarrow F_{Y_{i*}}(y) = \prod_{l \ne j} F_{V_{i,l}}(y) \overset{(c)}{=} y^{m_i - 1} \Rightarrow f_{Y_{i*}}(y) = (m_i - 1) y^{m_i - 2} \\
    &F_2(u) \overset{(d)}{=} \int_{0}^{1} \op{Pr}[V_{i,j} < u; (Y_k)^{1/w_k} < (y)^{1/w_i}, \forall k \ne i; V_{i,j} < y|Y_{i^{*}} = y]f_{Y_{i^{*}}}(y) dy \\
    &\overset{(e)}{=} (m_i - 1) \int_{0}^{u} y^{\mu_i - 1} dy + (m_i - 1) u \int_{u}^{1} y^{\mu_i - 2} dy = (m_i - 1) \left( \frac{u^{\mu_i}}{\mu_i} + u \frac{1 - u^{\mu_i - 1}}{\mu_i - 1} \right) \\
    &= \frac{m_i - 1}{\mu_i - 1} \left(u - \frac{u^{\mu_i}}{\mu_i} \right), \\
    &f_{U_{i,j}^*}(u) = \frac{dF_{U_{i,j}^*}(u)}{du} = \frac{dF_{1}(u)}{du} + \frac{dF_{2}(u)}{du} + \frac{dW(u)}{du} = \frac{m_i - 1}{\mu_i - 1}  + \frac{\mu_i - m_i}{\mu_i - 1} u^{\mu_i - 1} + P_{\bar{i}} \delta(u)\\
    &\op{UPI}_{i,j}^{(V)} = 2 \int_{0}^{1} u f_{U_{i,j}^*}(u) du = 2 \left(\frac{1}{2} \frac{m_i - 1}{\mu_i - 1} + \frac{\mu_i - m_i}{\mu_i - 1} \frac{1}{\mu_i + 1}\right) = \frac{m_i + 1} {\mu_i + 1}.
    \end{flalign*}
    Above we have (a), (d) from the law of total probability, (b), (c), (e) from properties of uniform variables and the MWS scheme.  Finally, it can be shown that $\op{UPI}_{i,j}^{(X)} = \op{UPI}_{i,j}^{(V)}$ and thus 
    \begin{flalign*}
    \op{UPI}_{i,j}^{(X)} = \frac{m_i + 1} {\mu_i + 1},
    \text{ where } \mu_i \triangleq \sum_{k=1}^G (m_k w_k)/w_i. \qed
    \end{flalign*}

\section{Proof of Theorem \ref{theorem:prob_non_ortho}} \label{app:theorem:prob_non_ortho}
From the MWS scheme (\ref{eqn:inter_cluster}), the probability group $i$ being selected is given by
    \begin{flalign*}
    P_i &= \op{Pr}[(Y_k)^{1/w_k} < (Y_i)^{1/w_i}, \forall k \ne i] \overset{(a)}{=} \int_{0}^{1} \op{Pr}[Y_k < y^{w_k/w_i}, \forall k \ne i | Y_i = y] f_{Y_i}(y) dy \\
    &\overset{(b)}{=} \int_{0}^{1} \prod_{k \ne i} \op{Pr}[Y_k < y^{w_k/w_i}] f_{Y_i}(y) dy = \int_{0}^{1} \prod_{k \ne i} F_{Y_k}(y^{w_k/w_i}) f_{Y_i}(y) dy \\
    &\overset{(c)}{=} \int_{0}^{1} y^{\sum_{k \ne i} m_k w_k/w_i } m_i y^{m_i - 1} dy = m_i \int_{0}^{1} y^{\mu_i - 1} dy = \frac{m_i}{\mu_i} = \frac{m_i w_i}{\sum_{k} m_k w_k}. \qed
    \end{flalign*}
Here (a) is from the law of total probability, (b) from the independence of $Y_k$, and (c) from the definition of $Y_i$ in (\ref{eqn:max_rep}), which leads to $F_{Y_i}(y) = y^{m_i}$ and $f_{Y_i}(y) = m_i y^{m_i-1}$.

\section{Proof of Theorem \ref{theorem:cdf_non_ortho}} \label{app:theorem:cdf_non_ortho}
The CDF for the SNR of user $j$ in group $i$ when group $i$ is selected is
\begin{flalign*}
  F_{S_{i,j}^{*}}(s) &= \op{Pr}[S_{i,j}^{*} < s] = \op{Pr}[S_{i,j} < s | \text{group i selected}] \\
  &= \underbrace{\op{Pr}[S_{i,j} < s; \text{group i selected}]}_{P_1} / \underbrace{\op{Pr}[\text{group i selected}]}_{P_i}\\
  P_1 &= \op{Pr}[S_{i,j} < s; Y_i^{1/w_i} > Y_k^{1/w_k}, \forall k \ne i] \\
  &= \underbrace{\op{Pr}[S_{i,j} < s; Y_i^{1/w_i} > Y_k^{1/w_k}, \forall k \ne i; Y_i \ne V_{i,j}]}_{P_2} \text{ ($V_{i,j}$ is not group $i$ representative)}\\
  &+ \underbrace{\op{Pr}[S_{i,j} < s; Y_i^{1/w_i} > Y_k^{1/w_k}, \forall k \ne i; Y_i = V_{i,j}]}_{P_3} \text{ ($V_{i,j}$ is group $i$ representative)}\\
  &P_2 \overset{(a)}{=} \sum_{l \ne j} \underbrace{\op{Pr}[S_{i,j} < s; V_{i,l}^{1/w_i} > Y_k^{1/w_k}, \forall k \ne i; V_{i,l} > V_{i,m}, \forall m \ne l]}_{P_4}\\
  &P_4 = \op{Pr}[S_{i,j} < s; V_{i,l} > Y_k^{w_i/w_k}, \forall k \ne i; V_{i,l} > V_{i,m}, \forall m \ne l,j; V_{i,l} > V_{i,j}] \\
  &\overset{(b)}{=} \int_{0}^{1} \op{Pr}[S_{i,j} < s; v > Y_k^{w_i/w_k}, \forall k \ne i; v > V_{i,m}, \forall m \ne l,j;v > V_{i,j}|V_{i,l}=v] f_{V_{i,l}}(v) dv \\
  &\overset{(c)}{=} \int_{0}^{1} \op{Pr}[S_{i,j} < s; V_{i,j} < v] \op{Pr}[Y_k^{w_i/w_k} < v, \forall k \ne i] \op{Pr}[V_{i,m} < v, \forall m \ne l,j] dv \\
  &\overset{(d)}{=} \int_{0}^{1} \op{Pr}[V_{i,j} < F_{S_{i,j}}(s); V_{i,j} < v] \underbrace{\prod_{k \ne i} \op{Pr}[Y_k < v^{w_k/w_i}]}_{v^{\sum_{k \ne i} \frac{m_k w_k}{w_i}}} \underbrace{\prod_{m \ne l,j} \op{Pr}[V_{i,m} < v]}_{v^{m_i - 2} = v^{\frac{m_i w_i}{w_i}} v^{-2}} dv \\
  &\overset{(e)}{=} \int_{0}^{F_{S_{i,j}}(s)} \op{Pr}[V_{i,j} < v] v^{\mu_i - 2} dv + \int_{F_{S_{i,j}}(s)}^{1} \op{Pr}[V_{i,j} < F_{S_{i,j}}(s)] v^{\mu_i - 2} dv\\
  &= \int_{0}^{F_{S_{i,j}}(s)} v v^{\mu_i - 2} dv + \int_{F_{S_{i,j}}(s)}^{1} F_{S_{i,j}}(s) v^{\mu_i - 2} dv\\
  &= \frac{\left[F_{S_{i,j}}(s)\right]^{\mu_i}}{\mu_i} + \frac{F_{S_{i,j}}(s)}{\mu_i - 1} \left(1 -  \left[F_{S_{i,j}}(s)\right]^{\mu_i-1}\right) = \frac{F_{S_{i,j}}(s)}{\mu_i - 1} - \frac{\left[F_{S_{i,j}}(s)\right]^{\mu_i}}{\mu_i (\mu_i - 1)}, \\
  P_2 &= (m_i - 1) P_4 = (m_i - 1) \left(\frac{F_{S_{i,j}}(s)}{\mu_i - 1} - \frac{\left[F_{S_{i,j}}(s)\right]^{\mu_i}}{\mu_i (\mu_i - 1)} \right).
\end{flalign*}
Above (a), (b) are from the law of total probability, (c), (d) from the independence of the user random variables and that $V_{i,l}$ is uniform, (e) from CDF of $Y_i$ (see App. \ref{app:theorem:prob_non_ortho}) and CDF of uniform variables $V_{i,m}$. Similarly for $P_3$, we have
\begin{flalign*}
  P_3 &\overset{(f)}{=} \op{Pr}[S_{i,j} < s; V_{i,j}^{w_k/w_i} > Y_k, \forall k \ne i; V_{i,j} > V_{i,l}, \forall l \ne j]\\
  &\overset{(g)}{=} \int_{0}^{1} \op{Pr}[V_{i,j} < F_{S_{i,j}}(s); v^{w_k/w_i} > Y_k, \forall k \ne i; v > V_{i,l}, \forall l \ne j|V_{i,j}=v]f_{V_{i,j}}(v) dv\\
  &\overset{(h)}{=} \int_{0}^{1} \op{Pr}[v < F_{S_{i,j}}(s)] \prod_{k \ne i} \op{Pr}[Y_k < v^{w_k/w_i}] \prod_{l \ne j} \op{Pr}[V_{i,l} < v] dv \\
  &\overset{(i)}{=} \int_{0}^{F_{S_{i,j}}(s)} v^{\sum_{k \ne i} \frac{m_k w_k}{w_i}} v^{m_i - 1} dv = \int_{0}^{F_{S_{i,j}}(s)} v^{\mu_i - 1} dv = \frac{\left[F_{S_{i,j}}(s)\right]^{\mu_i}}{\mu_i}.
\end{flalign*}
Above (f) is due to $V_{i,j} = Y_i$ being the representative for group $i$, (g) from the law of total probability, (h) from independence of random variables and that $V_{i,j}$ is uniform, and (i) from the CDF of $Y_i$ (see App. \ref{app:theorem:prob_non_ortho}) and CDF of uniform variables $V_{i,l}$.  From (\ref{eqn:prob_non_ortho}), $P_i = m_i / \mu_i$, we have
\begin{flalign*}
  F_{S_{i,j}^{*}}(s) &= (P_2+P_3)/P_i = \frac{\mu_i}{m_i} \left[(m_i - 1) \left(\frac{F_{S_{i,j}}(s)}{\mu_i - 1} - \frac{\left[F_{S_{i,j}}(s)\right]^{\mu_i}}{\mu_i (\mu_i - 1)} \right) + \frac{\left[F_{S_{i,j}}(s)\right]^{\mu_i}}{\mu_i}\right] \\
  &= \frac{\mu_i(m_i - 1)}{m_i(\mu_i - 1)} F_{S_{i,j}}(s) + \frac{\mu_i - m_i}{m_i (\mu_i - 1)} \left[F_{S_{i,j}}(s)\right]^{\mu_i}. \qed
\end{flalign*}

\ifCLASSOPTIONcaptionsoff
  \newpage
\fi



\bibliographystyle{IEEEtran}
%
%
\bibliography{./refs}



%




\end{document}